\documentclass[fleqn,usenatbib]{mnras}

\usepackage{newtxtext,newtxmath}

\usepackage[T1]{fontenc}

\DeclareRobustCommand{\VAN}[3]{#2}
\let\VANthebibliography\thebibliography
\def\thebibliography{\DeclareRobustCommand{\VAN}[3]{##3}\VANthebibliography}

\usepackage{graphicx}	
\usepackage{amsmath}	
\usepackage{subcaption}
\usepackage[utf8]{inputenc}
\usepackage[dvipsnames]{xcolor}
\usepackage{lastpage}
\usepackage{placeins}
\usepackage{capt-of}
\usepackage{caption}

\title[V486 Car]{The Close Binary  V486 Carinae}

\author[Erdem et al.]{Ahmet Erdem$^{1,2}$\thanks{E-mail: aerdem@comu.edu.tr},
Volkan Bak{\i}\c{s}$^{3}$,
Burcu \"{O}zkarde\c{s}$^{1,4}$,
Edwin Budding$^{5,6}$,
Mark G. Blackford$^{7}$,
\newauthor
Tom Love$^{8}$,
Michael D. Rhodes$^{9}$,
\& Timothy S. Banks$^{10,11}$
\\
$^{1}$Astrophysics Research Center \& Ulup{\i}nar Observatory, \c{C}anakkale Onsekiz Mart University, TR-17100, \c{C}anakkale, T\"{u}rkiye\\
$^{2}$Department of Physics, Faculty of Science, \c{C}anakkale Onsekiz Mart University, Terzio\u{g}lu Kamp\"{u}s\"{u}, TR-17100, \c{C}anakkale, T\"{u}rkiye\\
$^{3}$Department of Space Sciences and Technologies, Faculty of Sciences, Akdeniz University, 07058 Antalya, T\"{u}rkiye\\
$^{4}${Dept.\ of Space Sciences \& Technologies, Terzio\u{g}lu Kamp\"{u}s\"{u},} \c{C}anakkale Onsekiz Mart University, TR-17100, \c{C}anakkale, T\"{u}rkiye\\
$^{5}$Carter Observatory, 40 Salamanca Road, Kelburn, Wellington 6012, New Zealand\\
$^{6}$School of Chemical \& Physical Sciences, Victoria University of Wellington, PO Box 600, Wellington 6140, NZ\\
$^{7}$Variable Stars South, Congarinni Observatory, Congarinni, NSW, 2447, Australia\\
$^{8}$Centre for Astrophysics, University of Southern Queensland, Toowoomba, Australia\\
$^{9}$Brigham Young University, Provo, Utah 84602, USA\\
$^{10}$Department of Physical Science \& Engineering, William Rainey Harper College, 1200 W Algonquin Rd, Palatine, IL 60067, USA\\
$^{11}$Komatsu, 8770 W. Bryn Mawr Ave., Suite 100, Chicago, IL 60631, USA}

\date{Accepted XXX. Received YYY; in original form ZZZ}

\pubyear{\the\year{}}

\begin{document}

\label{firstpage}
\pagerange{\pageref{firstpage}--\pageref{lastpage}}
\maketitle

\begin{abstract}
The hitherto neglected close binary V486 Car is studied with the aid of newly applied satellite photometry (HIPPARCOS and TESS), high dispersion spectrometry (HERCULES) and ground-based B and V photometry. While the sinusoidal light variations are suggestive of a near-contact system, the stars have only shallow eclipse, so highly confident  parametrization becomes challenging. We find: $M_1 = 2.1 \pm 0.1$,  $M_2 = 0.4 \pm 0.1$; $R_1 = 3.20 \pm 0.02$, $R_2 = 1.48 \pm 0.01$; (${\odot}$); $T_{e1} = 10000 \pm 500$, $T_{e2} = 6200 \pm 200$ (K); distance = 162 $\pm$ 12 (pc). New times of minima for V486 Car have been examined, including recent observations from TESS. The role of the relatively significant O'Connell effect is examined. As well as the conspicuous asymmetry from the main effect of about 0.036 mag (V),
 a jitter, with amplitude of about 0.005 V mag and quasi-period
of order $\sim$ 10 d is noticed. There is a tendency for such photometric excursions at one maximum to precede those at the other.  As well, the O -- C data indicate the presence of a low mass star $\sim$0.3 M$_{\odot}$ in an orbit separated by a few AU from the close binary. More accurate and plentiful spectroscopic data would be requisite for further investigations. A brief discussion reviews possible approaches to understanding the system in the context of
near-contact binary scenarios.
\end{abstract}




\section{Introduction}
\label{sec:introduction}

The sixth magnitude periodic variable star V486 Carinae  (HD 84416, HIP 47951, WDS J09422-6655A) exhibits a smooth, sinusoidal, somewhat asymmetric light curve (LC) with the relatively short period of $\sim$1.094 d.  The apparent difference in light levels  of the two minima, while only $\sim$0.04 mag (V),    is about 30\% of the full amplitude of the variation. This has led to the light curves LCs being classified as of $\beta$  Lyr (EB) type \citep{Malkov_2006, Malkov_Oblak_2006}.  The form of the LC, together with the radial velocity results
(Section \ref{sec:spectroscopy}),  suggest  a close binary system as the essential cause of the photometric variability; the relatively  low amplitude pointing to shallow eclipses that we find herein. 

This hitherto neglected star was selected for further study, as part of our Southern Eclipsing Binaries Programme \citep{Idaczyk_2013}. 
\cite{Bakis_2024} discussed  paths for the evolution of very close and near-contact binaries (NCBs, \citeauthor{Yakut_2005}, \citeyear{Yakut_2005}), noting that various properties can be reconciled with Algol-like processes of mass, angular momentum and energy transfer between the components. This leads us to check if  such ideas can be substantiated  by observational evidence on V486 Car.

High-quality photometry, including data from the TESS \citep{Ricker_etal_2015} and Gaia \citep{Gaia_2016, Gaia_2022} sat ellites, are freely available as well as high resolution spectrometry. We can thus follow  \citeauthor{Russell_1912a}'s (\citeyear{Russell_1912a, Russell_1912b}) well-known approach to determine key absolute quantities.  Scenarios for the interactive evolution paths can then be considered.   The O'Connell effect \citep{OConnell_1951} is highly  relevant in this context. Its behaviour may offer insights into the extent of mass transfer and accretion structures.

Similar stable light-curve asymmetries have also been reported in other eclipsing binaries  (\cite{Yorulmaz_etal_2025} for DU Boo; \cite{Pribulla_etal_2011} for DU Boo and AG Vir).  The scale of the O'Connell effect can be reconciled
with other evidence favouring dynamic interaction
in the interpretation of the data.  There is a predictive element in such scenarios, particularly for more massive NCBs  that offers more testability than 
simple  {\it ad hoc} maculation. We consider this  further
in later sections of this paper.  See also 
\citep{Davidge_1984,Qian_2001,Liu_2003}.

The photometry section which follows is divided into  8 subsections. These deal with: (1)  ground-based B and V observations, (2) LC modeling using the  WinFitter (WF) program suite, (3) TESS photometry, (4) Times of Minimum light (ToMs), (5) the O'Connell effect, (6) Wilson-Devinney-based (WD+MC) fits to TESS data, (7) WF fits to TESS photometry, and (8) WD + MC fits to Congarinni and HIPPARCOS photometry. 

Section~\ref{sec:spectroscopy} discusses the spectroscopy. It deals with (1) data preparation and (2) determination of the optimal radial velocity (RV) semi-amplitudes. Combining the results of both these sections leads us to the absolute parameters, spelled out in Section~\ref{sec:absolute_parameters}.  Implications of these parameters regarding the origin, evolution and possible future condition of the system are presented in a final discussion and conclusions as section~\ref{discussion}.


\section{Photometry}
\label{sec:photometry} 


\subsection{B and V observations and preliminary analysis}
\label{sec:bv_observations}

BV photometry of V486 Car was carried out at the Congarinni Observatory (152° 52’ E, 30° 44’ S, Alt.\ 20 m) over 17 nights between January and April 2025 using a 51mm f4.9 refractor equipped with a ZWO ASI2600MM-P Mono CMOS camera and Johnson-Cousins BV filters. The field of view was 3.6 $\times$ 5.4 deg, which enabled the inclusion of a number of bright comparison stars. For data reduction, we used HD 84866 (V = 7.390, B -- V = $-0.041$) as the main comparison star. Primary and secondary extinction corrections were applied.

\begin{figure}
        \begin{center} 
          \includegraphics[width=1.0\columnwidth]{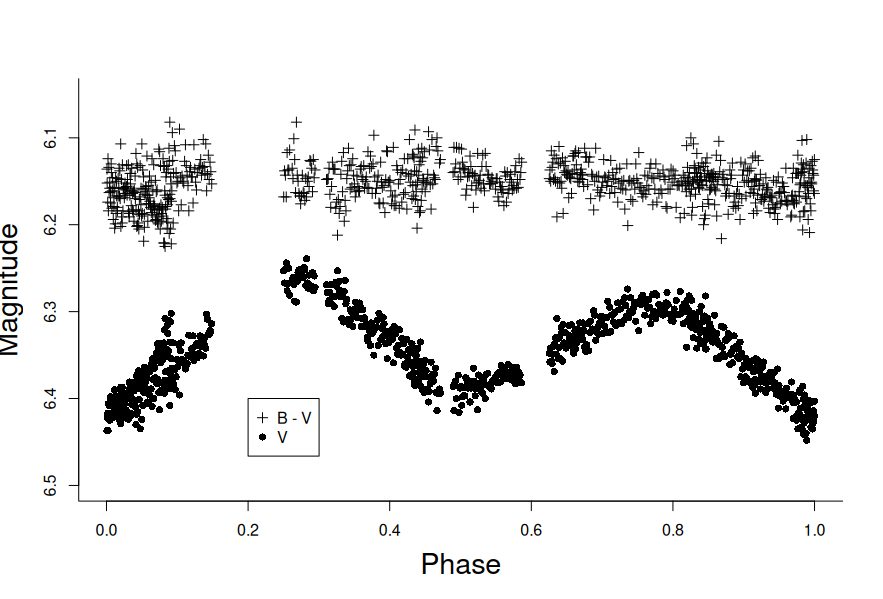}
        \end{center}
\caption{ Raw V data are plotted against orbital phase.  B -- V colours are offset by 6.1 mag from their actual values. Note the absence of significant colour (B -- C)  variation in this photometry.}
\label{fig:raw_photometry}
\end{figure}

The raw observations presented in Fig~\ref{fig:raw_photometry} form the essential  light curve in V, and the B -- V colour curve. Although the scatter is appreciable, the near constancy of the  B -- V colour trend over the phase cycle can be gathered. The average values of V and B -- V in these data is 6.337 and 0.053, respectively.  Using the Gaia DR3 estimate of $\sim$ 160 pc for the distance of V486 Car and a typical value of   interstellar extinction with the reddening ratio $R_V$ = 3.1 (cf.\ \citeauthor{Cardelli_1989}, \citeyear{Cardelli_1989}),  we derive an expected value of 0.052 for the observed colour, indicating the unreddened (B -- V)$_0$ of the target star to be $-0.001$.  Given the  scatter in the data of $\sim$0.01 mag, we can accept the photometry as confirming the A0 spectral type, or close to 10000 K effective temperature, of V486 Car \citep{Eker_etal_2018}.

\begin{figure}
        \begin{center} 
          \includegraphics[width=1.0\columnwidth]
          {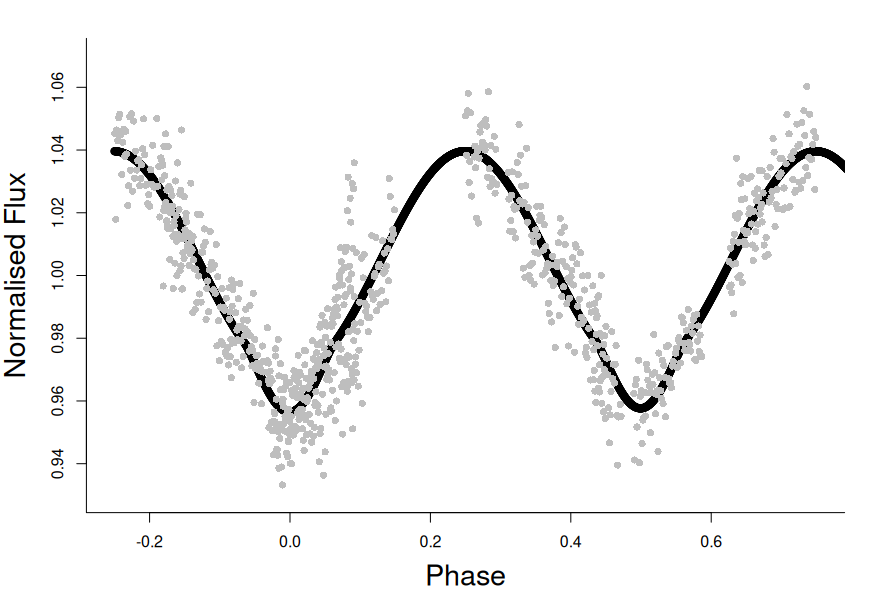}
           \label{fig:ightcurve2}
        \end{center}
        \hspace{-1cm}
\caption{Congarinni V data are plotted against orbital phase, together with a model LC (smooth curve) corresponding to the parameters given in Table~\ref{tab:winfitter_parameters}. }
\label{fig:mark_data_fit}
\end{figure}

Noting the relatively low information content of the photometry -- a quasi-sinusoidal LC without eclipses -- a simple, self-consistent preliminary model can be proposed involving two very similar Main Sequence components of $\sim$2.5 M$_{\odot}$ each, orbiting with the period of 1.094 d. Kepler's third law produces an orbital separation of 7.644 R$_{\odot}$. The mean fractional radii given in Table~\ref{tab:winfitter_parameters} then lead to stellar radii of $\sim$2.64 R$_{\odot}$. This is very close to the  mean radius expected for Main Sequence stars of the appropriate mass  ($\sim$2.65 R$_{\odot}$ given in Table~9.1 of \citeauthor{Budding_2022}, \citeyear{Budding_2022}), that would have photospheric effective temperatures of $\sim$10000 K. That table was based on the large range of high quality LC and RV data from eclipsing binary stars of \cite{Eker_etal_2018} .

Figure~\ref{fig:mark_data_fit} shows a binary star model fitting to the V photometry of the Congarinni Observatory after accounting for the O'Connell effect as an empirical superposition. To this end, we applied the WF program,\footnote{ See http://michaelrhodesbyu.weebly.com/astronomy.html for a comprehensive manual and to download the software.}   WF constructs  a physically  appropriate model that includes the effects of tidal and rotational distortion of the stellar envelopes (the {\em Radau} model, cf.\ \citeauthor{Kopal_1959}, \citeyear{Kopal_1959}, as well as light reflection effects). \cite{Banks_1990} presented a discussion of the method's optimisation procedure,  which follows the work of Bevington  (\citeyear{Bevington_1969}), particularly his Chapter 11.

\begin{table}
    \caption{Parameter values for an  initial  {\sc WF} model for the V light curve of V486 Car. The value of the normalized reference flux $U$ implies that the uneclipsed V reference magnitude is V = 6.271 mag. With the relatively large single-point error scale of 0.01, the fitting shows that a simple model that could explain the photometric data alone, on the basis of two very close near-identical A0-type MS stars at orbital inclination  $\sim$52\degr\, together with a hot region centered around the phase of the first maximum light, is plausible.}
    \begin{tabular}{l|l|l}
    \hline
    Parameter                           & Value                     & Uncertainty  \\
    \hline
    Reference light level $U$                  & 0.9807                    & 0.0019\\
    Fractional luminosities $L_1, L_2$   & 0.502, 0.498             & 0.02, 0.02\\        
    Relative radii $r_1, r_2$           & 0.345, 0.351              & 0.004, 0.004 \\   
    Orbit Inclination $i$               & 52$^{\circ}$              & 0.11$^{\circ}$ \\
    Mass ratio $ q $                    & 1.0                       & (ex hypothesis)\\
    Phase correction $\Delta \phi_0$    & 0.0008                    & 0.0011 \\
    Data scatter $\sigma$      & 0.01                      & --- \\
    $\chi^2/\nu$                        & 1.27                      &  --- \\
    \hline
    \end{tabular}
    \label{tab:winfitter_parameters}
\end{table}

In  Table~\ref{tab:winfitter_parameters}, we present a set of parameters  corresponding to the photometric model illustrated in Fig~\ref{fig:mark_data_fit} . This model,  mentioned above and involving two near-identical A0 dwarf stars, was influenced by its self-consistency and agreement with the Gaia parallax. 
The main parameters are listed in  Table~\ref{tab:winfitter_parameters}, where the fractional luminosities $L_1$, $L_2$ are normalized so that their sum is unity;  $r_1, r_2$ are mean radii of the two stars, divided by the semi-major axis of the relative orbit. Uncertainties, associated with the adjustable parameters are derived from numerically inverting the curvature Hessian in the vicinity of the $\chi^2$ minimum (\citeauthor{Bevington_1969}, \citeyear{Bevington_1969}, Ch,\ 11). This Hessian should be positive definite for a properly posed data-modeling problem.  The resulting parameter uncertainties  then include the effects of inter-correlations between the {optimised parameters}.

The self-consistency of this Main Sequence model with photometric data alone leads us to an important issue in LC analysis, namely, non-uniqueness in model parametrization, or the ambiguity that attends solutions of LCs produced by least-squares fitting techniques. Section~\ref{sec:spectroscopy} introduces the critical role of spectroscopy in restricting alternative models.

\subsection{TESS photometry}
\label{sec:tess_obs}

\begin{table}
    \centering
    \caption{Observation Summary by TESS Sector and Cadence}
    \label{tab:TESS_sectors}
    \label{tab:observations}
    \begin{tabular}{|c|r|l|}
        \hline
        Sector & Cadence  & Observation Dates \\
        \hline
        09              & 120-s             & February 28 to March 25, 2019 \\
        10              & 120-s             & March 26 to April 22, 2019 \\
        11              & 1800-s            & April 23 to May 20, 2019 \\
        36              & 600-s             & March 07 to April 01, 2021 \\
        37              & 600-s             & April 02 to April 28, 2021 \\
        38              & 600-s             & April 29 to May 26, 2021 \\
        63              & 120-s             & March 10 to April 06, 2023 \\
        64              & 120-s             & April 06 to May 03, 2023 \\
        65              & 120-s             & May 04 to June 01, 2023 \\
        90              & 200-s             & March 12 to April 09, 2025 \\
        \hline
    \end{tabular}
\end{table}

We downloaded the TESS data of V486 Car (TIC 370960295) from the Mikulski Archive for Space Telescopes (MAST)\footnote{https://mast.stsci.edu/portal/Mashup/Clients/Mast/Portal.html} \citep[cf.][]{JEN16}.  According to the archive, V486 Car was one of the   more frequently observed objects in the TESS programme: i.e.\ in 10 sectors from 2019 to 2025 (Table~\ref{tab:TESS_sectors}). 

Simple Aperture Photometry (SAP) measurements were used for the present work, since the Pre-search Data Conditioning Simple Aperture Photometry (PDCSAP) detrending introduces artificial side-effects, associated with the search for planetary transits.

\subsection{Times of minimum light (ToMs)}
\label{ToMs}

\begin{figure}
\centering
    \includegraphics[scale=0.50]{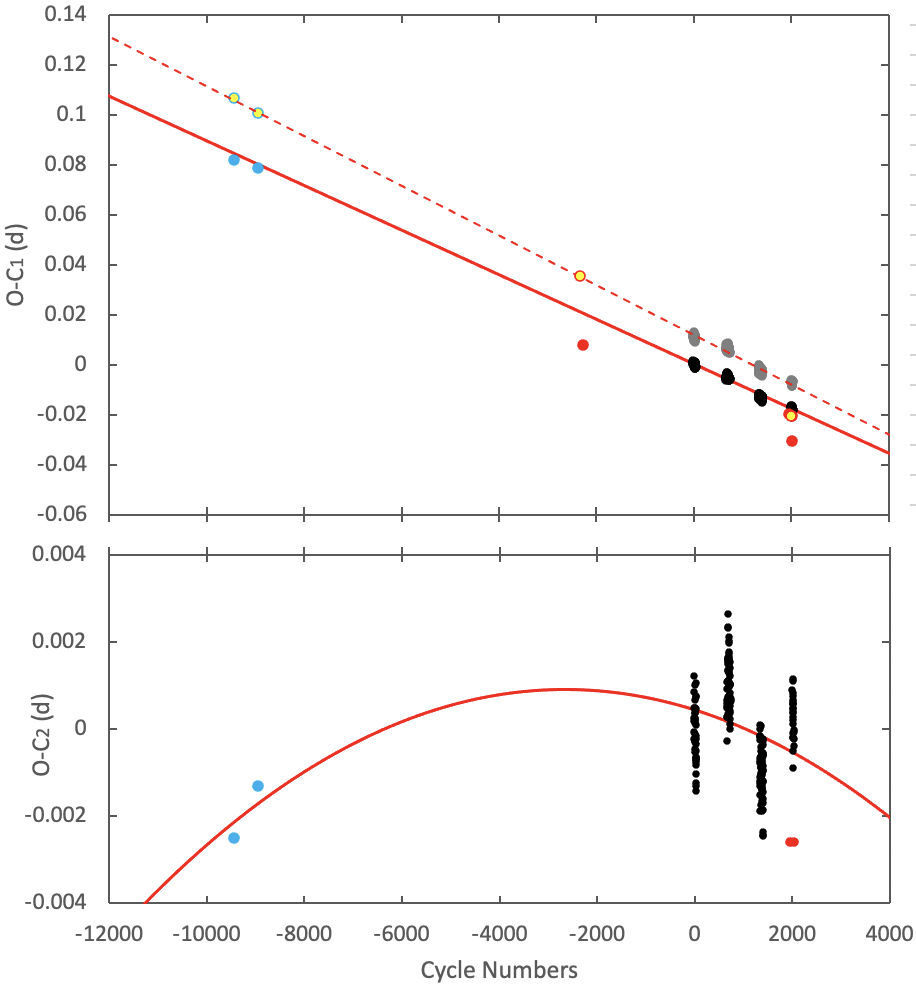}
    \caption{O-C trends of V486 Car. Top panel: Linear fit. Bottom panel: quadratic fit to residuals of linear fit. The blue and blue-yellow circles represent the Hipparcos primary and secondary ToMs, the red and red-yellow circles represent our BVI LCs' ToMs, and the black and gray circles represent the TESS ToMs.} 
    \label{fig:o-c_all}
\end{figure}

There are no previous minima times because V486 Car is bright, which complicates ground-based observations.
Photometric observations of the system were made by the Hipparcos \citep{Perryman_1997} and TESS \cite{Ricker_etal_2015} satellites. This paper includes the first reported ground-based observations to time light minima  V486 Car (Fig.~\ref{fig:o-c_all}).

Because the Hipparcos observations are distributed over three years from late 1989 to early 1993, a method similar to that of \cite{Zasche_etal_2014} was applied to estimate appropriate ToMs.  This yielded two primary and two secondary ToMs. Four primary and three secondary ToMs were also obtained from the ground-based BV observations presented here. These ToMs are collected in Table \ref{table:ToMs} in the appendix. A further 207 primary and 203 secondary ToMs were obtained from the intense TESS coverage. The well-known \cite{Kwee_1956} method and polynomial fittings were employed to derive ToMs from both the BV and TESS photometry.

Conventional O--C (observed minus calculated ToM) analysis was performed to check on the orbital period of V486 Car. A primary ToM ($T_0$ = 2458559.1579 BJD) was selected from the TESS data to calculate the O--C values, 
 with the orbital period ($P$ = 1.093893 d) taken from the Hipparcos catalog. The (O--C)s are plotted against  cycle numbers in Fig.~\ref{fig:o-c_all}.

Looking at Fig.~\ref{fig:o-c_all}, we can see that the (O--C)s show a first-order linear trend with decreasing slope. Somewhat surprisingly, the (O--C)s of the secondary ToMs show a distinctly separate linear distribution with a shift of approximately 0.01 days across all data. This is probably because the asymmetry created by the O'Connell effect in the photometric LCs (Hipparcos, TESS, and BVI). Starting from Min II (phase 0.5) we find the Max II light levels to be relatively low, thus shifting the LCs' midpoints of Min II  to the right (toward advancing phase). Therefore, to study properly  orbital period effects for V486 Car, the primary ToMs need to be  first taken into account.  The following equation was derived for the updated ephemeris, using the least squares method:
\begin{equation}
T_{\rm min} \, {\rm (BJD)}  = 2458559.1582(1) + 1.0938841(1) \times E
\end{equation}
Here, and in Eq. (2) below, the numbers given in parentheses represent the errors in the last digit(s) of the quoted values.
If we take into account the two primary ToMs estimated from the Hipparcos data, we can construct a second-order polynomial model, as shown in the bottom panel of Fig. \ref{fig:o-c_all} using the primary ToMs. The quadratic equation representing this model is as follows:

\begin{eqnarray}
\label{eq:quadratic}
T_{\rm min} \, {\rm (BJD)} 
& = & 2458559.1586(1) + 1.0938837(1) \times E \nonumber \\
 &  & - 6.63(1.26) \times 10^{-11 } \times E^2 \nonumber \\
\end{eqnarray}
The second-order term in Eq. \ref{eq:quadratic} indicates that the orbital period of V486 Car is decreasing at the slow rate of about 
0.0038 $\pm$ 0.0007 s per year  over the last $\sim$ 35 year interval.  This decrease in the orbital period may have occurred as a result of mass transfer between the components, or perhaps matter lost from the system via stellar winds.  

Using Kepler's third law to the mass-transferring system, with conservation of total mass and angular momentum, the following well-known formula is derived: 
\begin{equation}
\label{eq:period_change}
\frac{\dot{P}}{P} = 3 \left( \frac{M_{loser}-M_{gainer}}{M_{loser} M_{gainer}} \right)\dot{M}_{loser}
\end{equation} 

Taking the mass of the primary to be greater than that of the secondary, Eq. \ref{eq:period_change}  produces a period decrease. Substituting the mass values from Table \ref{table:abs_par} and the period decrease rate calculated from the quadratic O--C change, we find the rate of mass transfer to be $(5.7 \pm1.3) \times 10^{-9}$ M$_{\odot}$ yr$^{-1}$. 
The uncertainty is dominated by the error in $dP/dt$, with a smaller contribution from the uncertainty in $M_2$. 

\begin{figure}
\centering
    \includegraphics[scale=0.45]{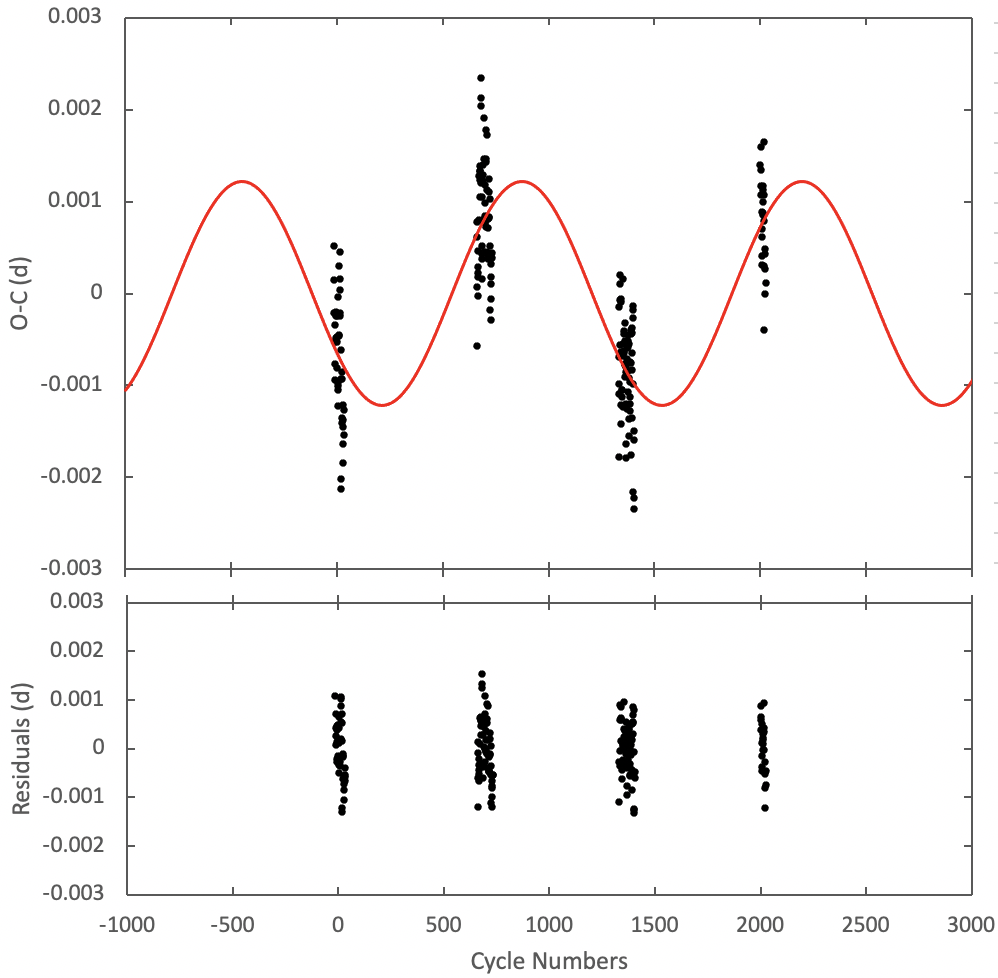}
    \caption{O--C change of the TESS primary ToMs of V486 Car. Top panel: Sinusoidal fit. Bottom panel: residuals. } 
    \label{fig:lite}
\end{figure}

As well, an alternating variation of the (O--C)s of TESS primary ToMs is noticeable in the bottom panel of Fig.~\ref{fig:o-c_all}. We fitted a sinusoidal model to these data and compared this model with the observational data in Fig.~\ref{fig:lite}. According to these results, the (O--C)s of the TESS data include a cyclical variation with a period of $3.96 \pm0.06$ years and an amplitude of $0.0012 \pm0.0003$ days. A similar sinusoidal variation is observed in the (O--C)s of the TESS secondary ToMs.
However, due to the limited temporal coverage, consisting of four relatively short observing intervals, the (O--C) diagram may be affected by aliasing. Therefore, although the adopted solution provides a satisfactory fit to the data, alternative periodicities -- particularly at shorter periods -- cannot be ruled out. Future observations with improved time coverage are required to uniquely constrain the true period. 

One interpretation of the cyclic O--C behaviour in Fig.~\ref{fig:lite} is the light-time effect (LTE)  related to the presence of an   orbiting  third body \citep{Irwin_1959}. The  fitting of a light travel-time effect (LTE) to the cyclic O--Cs indicates V486 Car moves in a circular orbit around the common center of mass of  a triple system. Applying Kepler's third law to this system (third body + V486 Car) allows us to calculate the mass of the third body and its angular distance from V486 Car. Accordingly, depending on the inclination of the axis  of the  circular orbit of V486 Car so that adopted for the axis of the wide triple system, $i_{12}$, we find: 
\begin{enumerate}
    \item{$M_3$ = 0.16 M$_{\odot}$ for $i_{12}=90\degr$,}
    \item{$M_3$ = 0.18 M$_{\odot}$ for $i_{12}=60\degr$,}
    \item{$M_3$ = 0.33 M$_{\odot}$ for $i_{12}=30\degr$.} 
\end{enumerate}
The angular separation of the third body from the inner binary is estimated to be 21.33 mas, assuming an orbital inclination of $i_{12}=30\degr$.

\subsection{The O'Connell effect}
\label{O_Connell}

V486 Car shows an appreciable asymmetry in its LCs. We have taken this to be an `O'Connell effect' \citep{OConnell_1951}.  This is generally regarded  as a difference in brightness ($I$) of the two
shoulder regions, i.e.\,  those parts of the LC  outside of the two minima.  The primary minimum is the deeper one, and the following high light level is the first maximum ($max_1$).  $max_2$ is the immediately following maximum. The O'Connell measure,
in magnitudes,
$\Delta m$ is then $m_{\rm max_2} - m_{\rm max_1}$.

In the case of V486 Car, $\Delta m$ was always found to be positive, so the first shoulder is the brighter. In Fig.~\ref{fig:raw_photometry} this brightness excess is, on average, about 0.036 V magnitudes.  In the TESS  LC shown as Fig.~\ref{fig:wd_hotspot_model}, this falls to about 0.017 mag.  The different scales of the O'Connell effect between TESS and ground-based data are suggestive of a locally heated region, perhaps indicative of  mass and energy transfer.

\begin{figure}
\centering
    \includegraphics[scale=0.38]{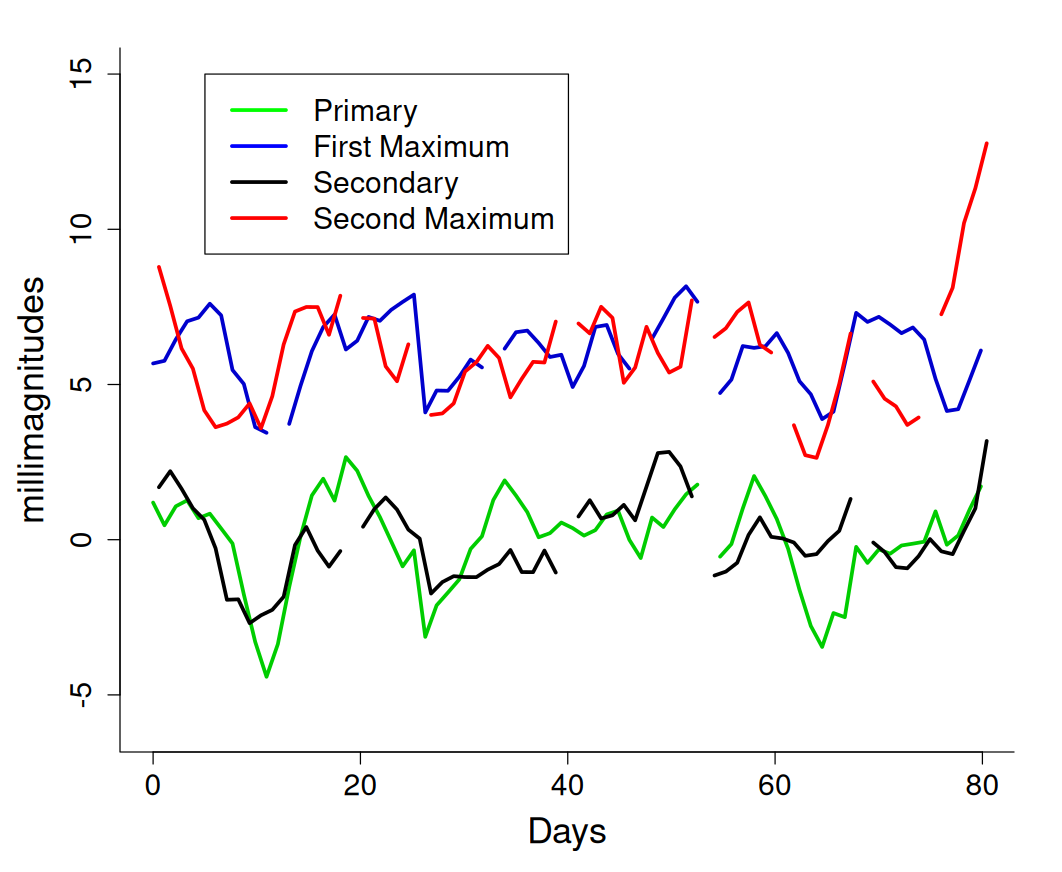}
    \caption{TESS photometry of V486 Car reveals quasi-periodic jitter in the LCs associated with fine structure in the O'Connell effect. The plots show variation of minima and maxima magnitudes from their average values. $m_{\rm max_2}$ and $m_{\rm max_1}$ values offset by 6 millimag.
    \label{figureZZZ}}
\end{figure}

We may summarize the salient properties of this O'Connell effect through the following main points:  
\begin{enumerate}
    \item{There is a fairly repetitive, positive basic  O'Connell effect  of about 0.036 mag (V) or 
0.017 mag at the effective wavelength of the TESS) photometer.}
    \item{A jitter, with amplitude of about 0.005 V mag and quasi period of order $\sim$ 10 d appears on top of the basic O'Connell effect.}
    \item{This jitter affects, and can be measured at, both maxima and both minima; i.e.\ it affects the whole system.}
    \item{There is a correlation between the various   jitter effects, measured at different phases,  
     with a tendency for excursions at one maximum to precede those at the other
     (see Fig.~\ref{figureZZZ}).}
\end{enumerate}

\subsection{WD+MC fits to TESS LCs}
\label{sec:wd+mc fits}

The \citeauthor{Wilson_Devinney_1971} (\citeyear{Wilson_Devinney_1971}, WD) numerical technique  models the LC of a  binary star by taking into account the effects of tidal distortion and radiative interaction, regarding the surfaces of the components as Roche equipotentials (see Ch. 3  of \citeauthor{Kopal_1959}, \citeyear{Kopal_1959}). This representation becomes significant in situations of relatively close pairs, particularly those in a near-contact arrangement, as becomes apparent for V486 Car.

The original WD program  has been combined with a `Monte Carlo' (MC) optimisation procedure, as discussed in \citet{Zola_etal_2004}. In this WD+MC method, an input range is defined for the adjustable parameters, and a near-minimum  {$ \chi^2 $} region is generated accordingly. This is thoroughly searched in repeated iterations to map out the location of optimal parametrizations (`best-fits'). The size of the surrounding region of near-optimal parameter hyperspace, which relates to their uncertainties, depends on the number of free parameters in the search. The uncertainties  tend to increase with the number of free parameters. 
The Monte Carlo search procedure provides confidence intervals at different levels (68\%, 90\%, and 98\%), enabling a direct evaluation of parameter uncertainties. In this application, we adopt the 90\% confidence intervals for the error estimation of the fitted parameters, which offers a conservative assessment of the uncertainties.

In our WD+MC procedure, the effective temperature ($T_1$) of the primary star is usually taken as a known and fixed parameter.  The spectral type of V486 Car was assigned as A0V by \cite{Houk_Cowley_1975}, as confirmed  from the derived colour in Section~\ref{sec:bv_observations}. 
As well, the 2MASS J--K colour of $0.101 \pm0.026$ mag \citep{Cutri_etal_2003} provides an additional consistency check for an early-A classification of the system, although this infrared colour refers to the combined light of the binary and may still be slightly affected by interstellar reddening. 
From the calibration data  of \cite{Eker_etal_2018}, the corresponding effective temperature  $T_1$ = 10000~K, was adopted in our LC analysis. The effective temperature of the secondary ($T_2$) was adjustable in the range of 5000 to 10000 K.  A quadratic limb-darkening law was assumed; the coefficients being taken from \citet{Claret_2017} according to the effective temperature and  filter used. The bolometric gravity darkening exponent ($g_1$) and albedo ($A_1$) of the primary component were adopted, following regular WD procedure, to be 1.0, assuming that the primary has a radiative atmosphere \citep{vonZeipel_1924,Rucinski_1969}. If the effective temperature ($T_2$) of the secondary becomes larger than 7200 K in the WD iterations, $g_2$ and $A_2$ are taken as 1.0, otherwise, with $T_2 < 7200$ K, $g_2 = 0.32$ and $A_2 = 0.5$, assuming that the secondary has a convective atmosphere \citep{Lucy_1967,Rucinski_1969}.  

The input range of the orbital inclination was set to $30^\circ < i < 70^\circ$, taking into account the {\sc WF} LC fittings in Section~\ref{sec:bv_observations}.   Possible changes in $T_0$ and $P$ relate to the input range for phase-shift $\Delta \phi$. This allows for a zero-point displacement in the phases, normalized to the orbital period. This range was set to $-0.01 < \Delta \phi <0.01$.  Based on the RV results in Table~\ref{tab:orbit_pars}, that will be discussed in Section~\ref{sec:spectroscopy}, the mass ratio ($q$) was set at  0.174. The input range of the surface potentials of both components ($\Omega_{1,2}$) was set to 2.0 -- 4.0. 
A range from 0.50 to 0.95 was entered for the fractional luminosity of the primary component ($L_1$). 

In the SAP data  of all sectors the CROWDSAP parameter, i.e.\ the ratio of target to total flux in the photometric aperture, is set to between 0.03 -- 0.04. The input range for the relative contribution of third light ($l_3$) was therefore set to 0.01 -- 0.10.

For the `hot spot' model (to be discussed below), initially, the third light parameter ($l_3$) was  adjustable. However, since its uncertainty was of the same order as its value, third light was not included in the final LC fittings.

In order to deal with the relatively large O'Connell effect that appears at almost the the same phase in all the  LCs,  both cool and hot spot alternatives  were added to the fitting options. The hot spot was taken to be on the secondary component; it being associated with mass transfer from the primary. 

Considering that the effective temperature of the secondary would be low enough to allow it to have a convective envelope, the cool spot was also placed on the secondary component. Assuming mass transfer to be essentially in  the  equatorial plane, the hot spot colatitude was taken as $\beta = 90\degr$, while a input range of $0^\circ < \beta < 180^\circ$ was set for the cool spot latitude.
An input range of $0\degr<\lambda<360\degr$ was set for the spot longitude, $10\degr<\gamma<60\degr$ was set for the spot angular radius and $0.50<\kappa< 1.50$ for the spot's increased temperature factor.

\begin{figure}
\centering
    \includegraphics[scale=0.45]{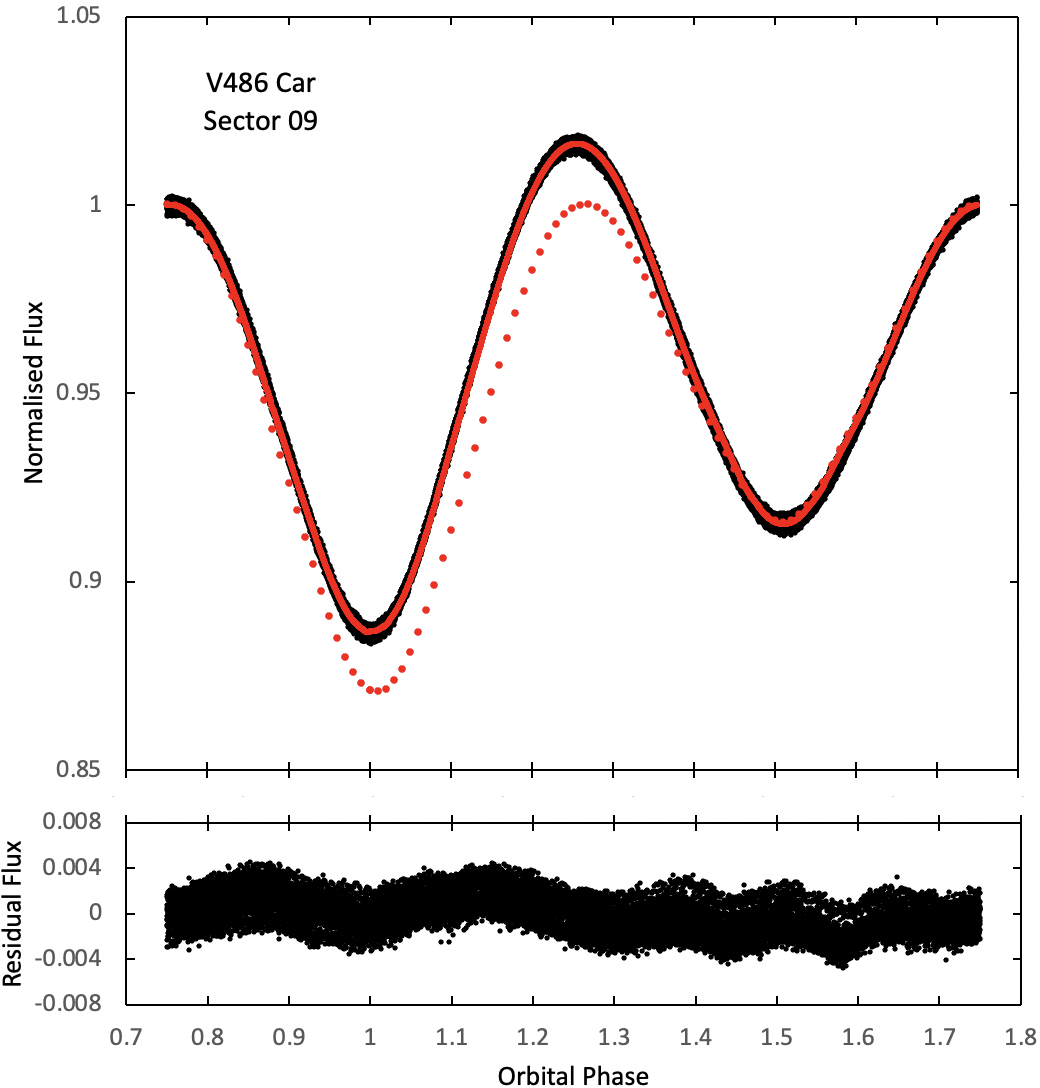}
    \includegraphics[scale=0.15]{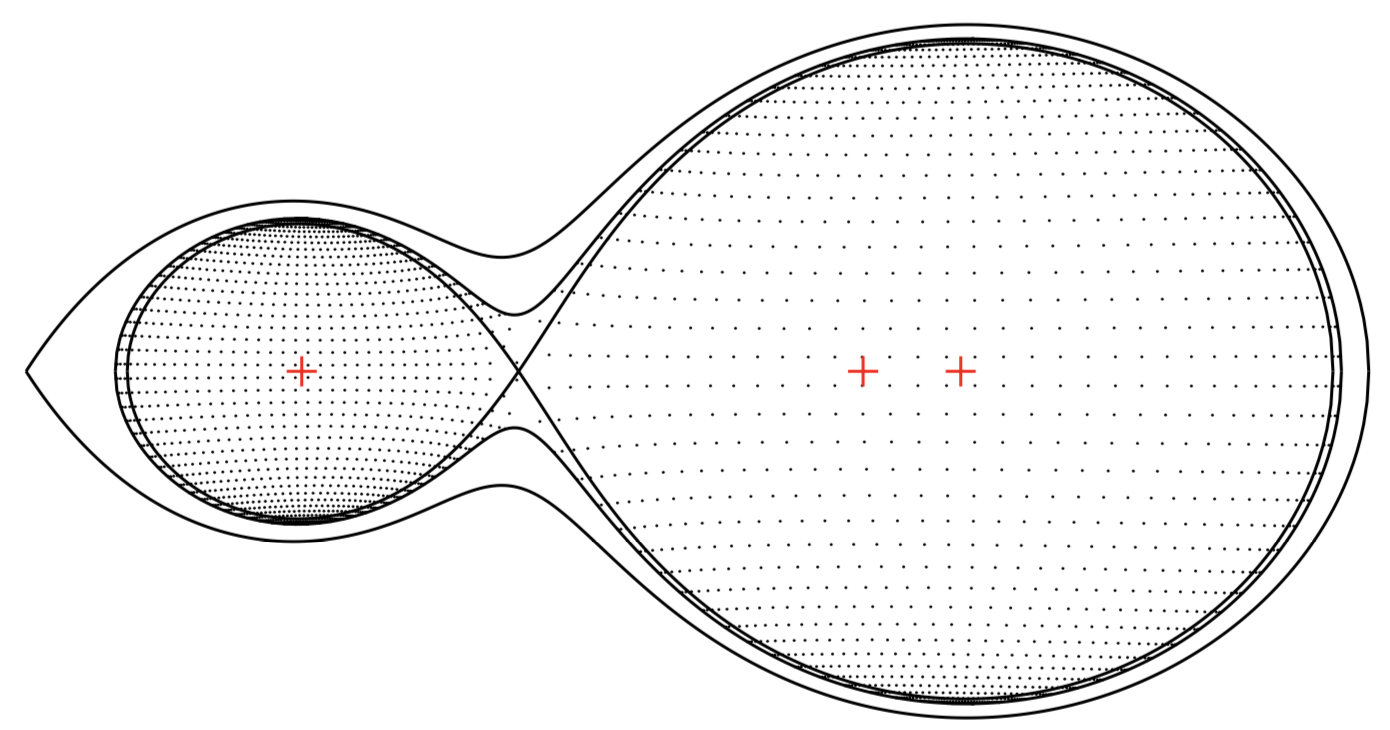}
    \includegraphics[scale=0.15]{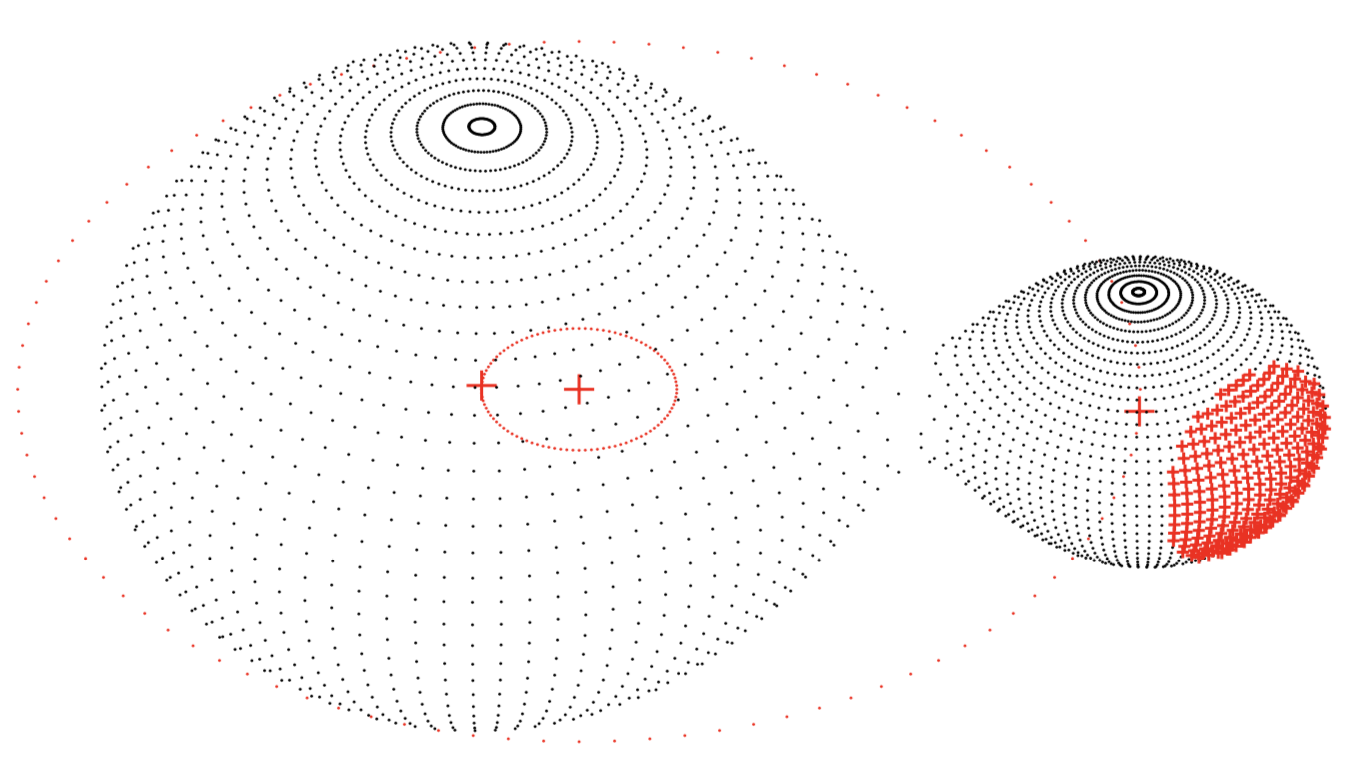}
    \caption{
    Top panel: TESS LCs in sector 09 with the WD+MC hot spot model fitting. 
    The red dotted light curve shows the theoretical LC without spots. 
    Residuals to the LC model are plotted in the lower figure. 
    Bottom panel: Critical Lagrangian surfaces and model of V486 Car (left), and 3D representation of V486 Car at phase 0.25 with a hot spot (right). } 
    \label{fig:wd_hotspot_model}
\end{figure}

\begin{figure}
\centering
    \includegraphics[scale=0.45]{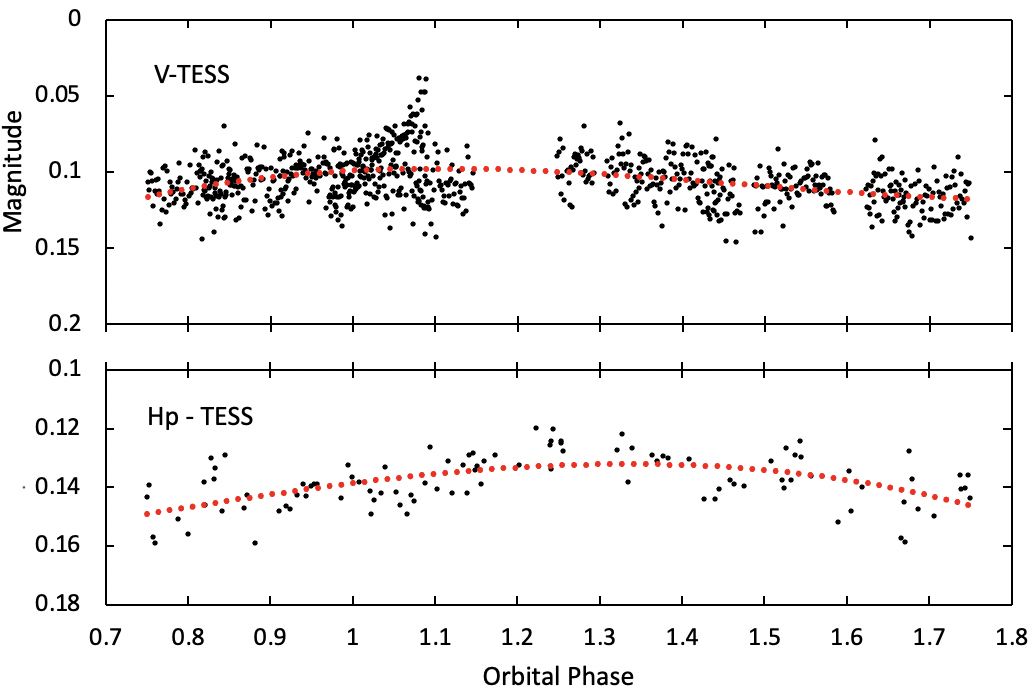}
    \includegraphics[scale=0.45]{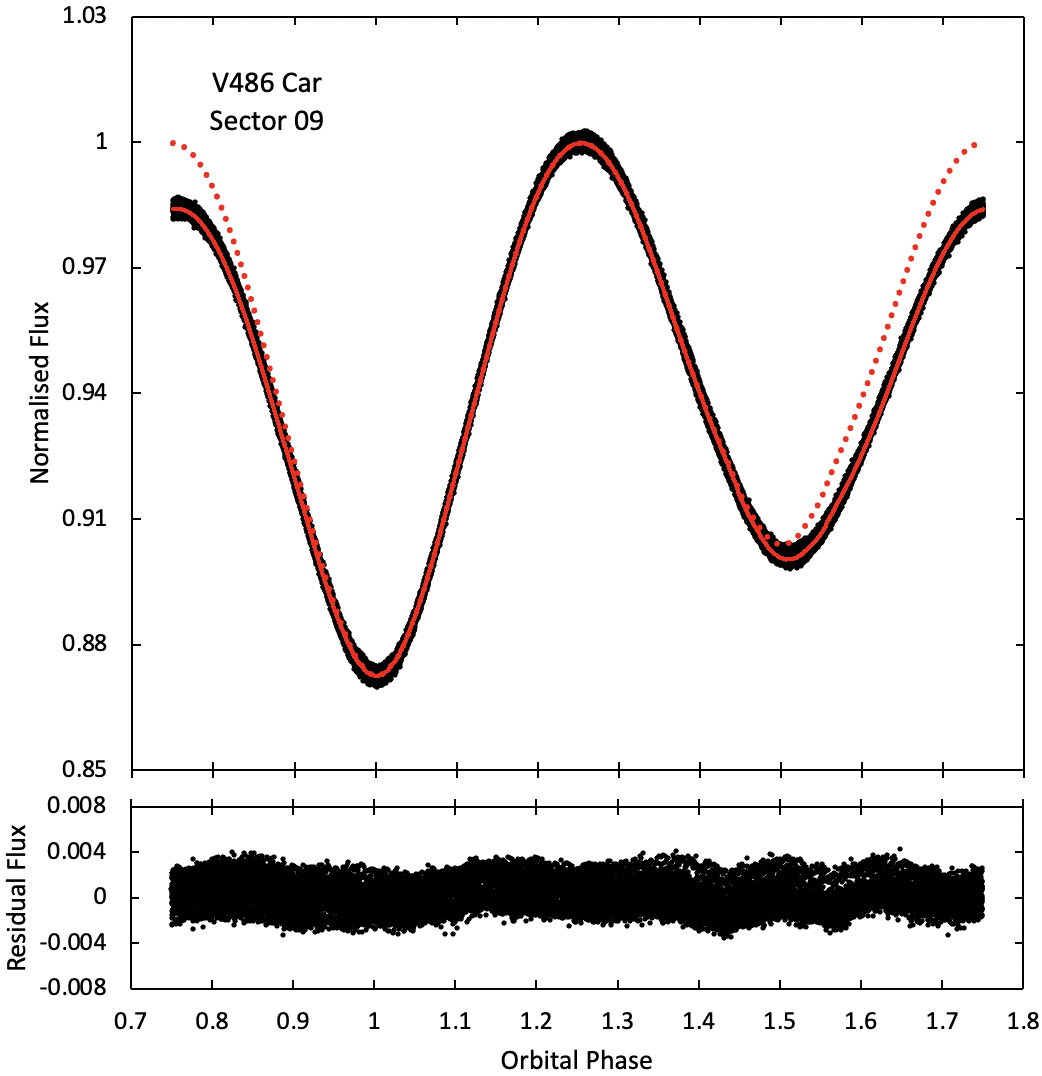}
    \includegraphics[scale=0.15]{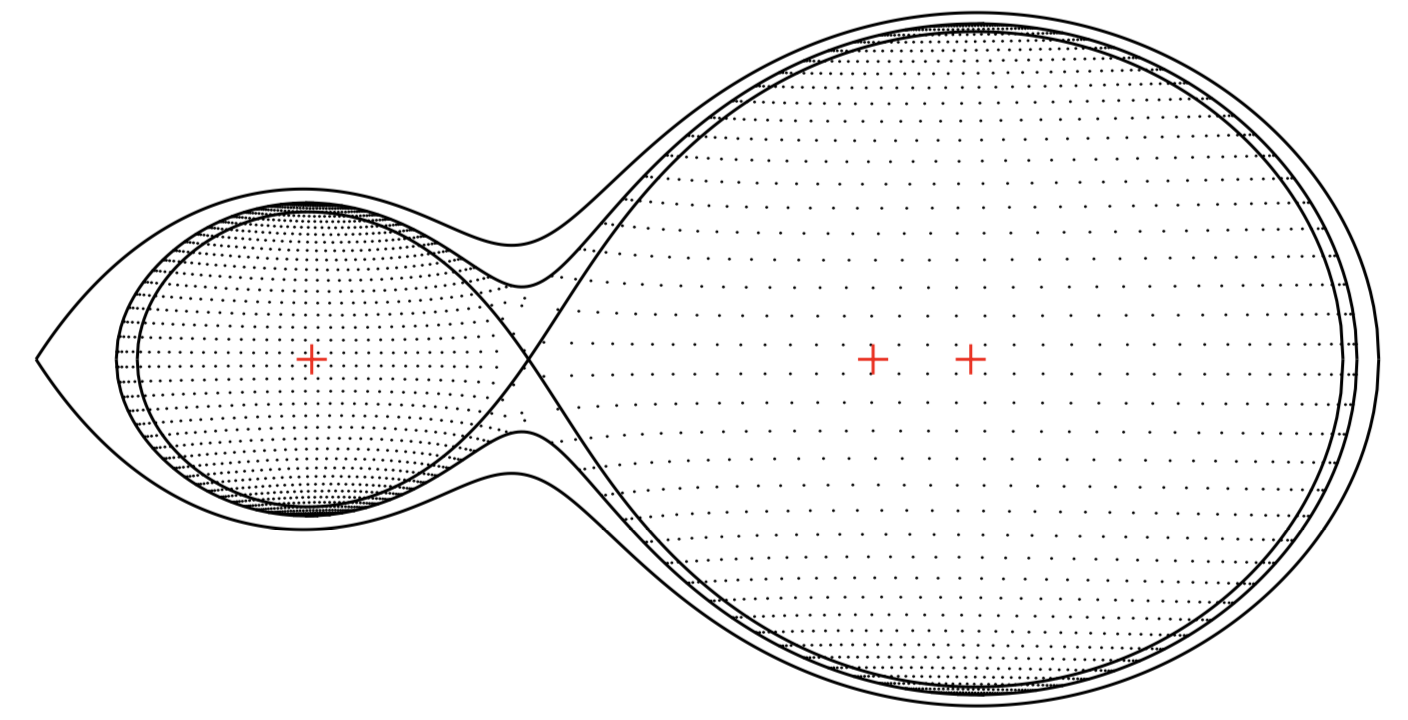}
    \includegraphics[scale=0.15]{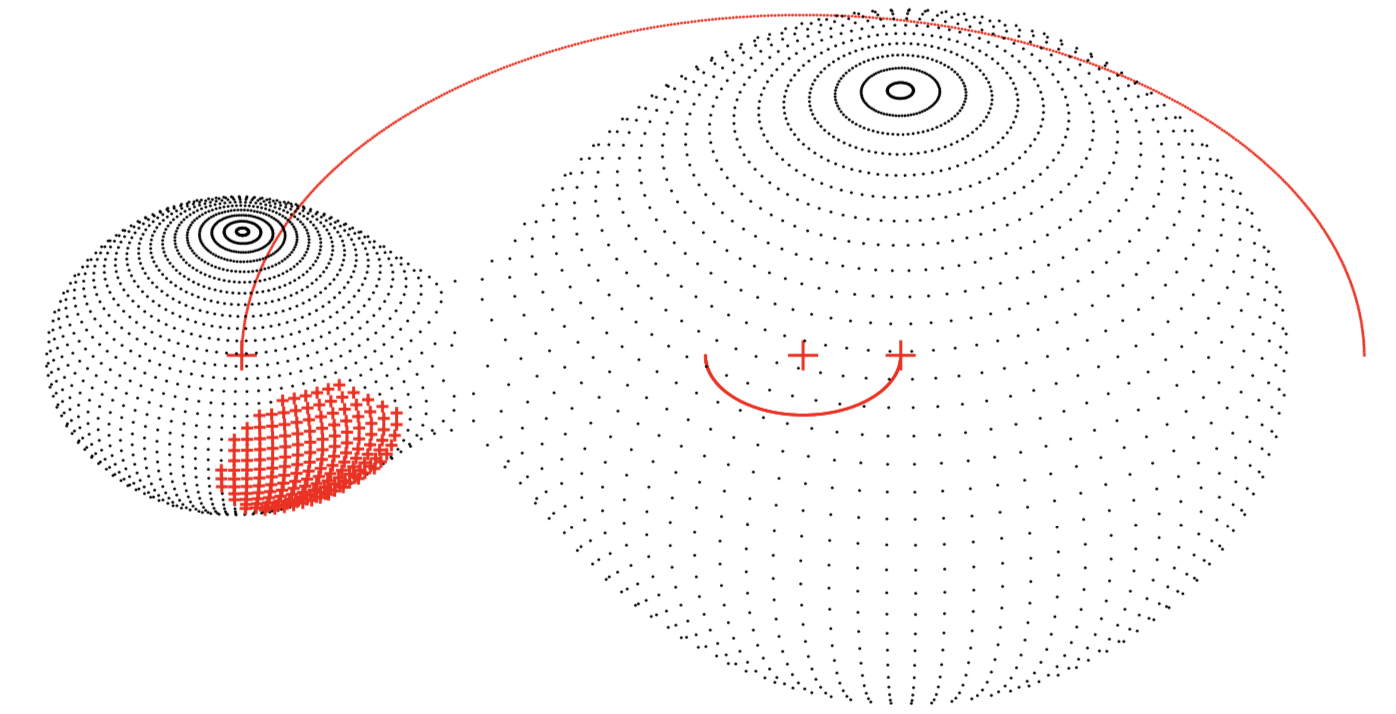}
    \caption{
    Top panel: V -- TESS and Hp -- TESS colour curves of V486 Car. The red dots represent the trend of colour curves. 
    Middle panel: TESS LCs in sector 09 with the WD+MC cool spot model fitting. 
    The red dotted light curve shows the theoretical LC without spots. 
    Residuals to the LC model are plotted in the lower figure. Bottom panel: Critical Lagrangian surfaces and model of V486 Car (left), and 3D representation of V486 Car at phase 0.75 with a cool spot (right). } 
    \label{fig:wd_coolspot_model}
\end{figure}

The adopted parameter values for the hot and cool spot models for each sector's LC are given in Tables \ref{Tab:WD+MC_TESS_hotspot_Model} and \ref{Tab:WD+MC_TESS_coolspot_Model}, respectively, while the weighted average and weighted error values obtained from these separate sector LC solutions for each model are presented in Table \ref{Tab:WD+MC_TESS Models}. For the hot spot approach, the mean values from the individual sector solutions are a longitude of $142^\circ$, an angular radius of $42^\circ$, and a temperature factor of $1.33$, with standard deviations of $1^\circ$, $5^\circ$, and $0.05$, respectively.
For the cool spot approach, the mean values are $\beta = 100^\circ$, $\lambda = 295^\circ$, $\gamma = 34^\circ$, and $\kappa = 0.61$, with standard deviations of $2^\circ$, $1^\circ$, $1^\circ$, and $0.04$, respectively. Comparisons of the photometric observations with the hot and cool spot models are also shown in Figs. \ref{fig:wd_hotspot_model} and \ref{fig:wd_coolspot_model}, respectively.
The 3D projected illustrations of V486 Car, using the program {\sc Binary Maker} \citep{Bradsteet_Steelman_2002}, are presented in the bottom panels of these figures.

Note that the colour curves  can also provide  clues as to whether the O'Connell effect is due to a hot  or  cool spot. The B -- V color curve, derived from almost simultaneous B and V observations of the system, is shown in Fig. \ref{fig:raw_photometry}. From that it is not possible to draw a definite conclusion, due to the observational scatter. However, this effect can be taken to manifest itself at  the same  phase for $\sim$30 years, Therefore, although the ground-based and satellite observations were not simultaneous, colour curves can be constructed from V, Hipparcos and TESS data at the same   phases, as presented in the top panel of Fig. \ref{fig:wd_coolspot_model}. The increased reddening towards phase 0.75 points to cool maculation becoming effective around this phase.

In Table \ref{Tab:WD+MC_TESS Models}, the fill-out factor $f$ is defined by \cite{Lucy_Wilson_1979} to measure the degree of contact of a component star within a binary system as follows
\begin{equation}
\label{eq:fill-out}
f = \frac{\Omega_{inner}-\Omega}{\Omega_{inner}-\Omega_{outer}},
\end{equation} 
where $\Omega$, $\Omega_{inner}$ and $\Omega_{outer}$ are the potentials of the common photosphere and of the inner and outer critical Lagrangian surfaces.
The fill-out factor $f$ for an over-contact system will lie between $0 \leq f \leq 1$ i.e., from  contact with the inner critical surface ($f = 0$) to contact with the outer one ($f = 1$). 

\begin{table}
\begin{center}
\caption{Results of the WD+MC fitting to the TESS light curves of V486 Car. $r$ (volume) is the radius of a sphere having the same volume as the tidally distorted star. $l_3$ is the third light contribution to the total light at phase 0.25.
\label{Tab:WD+MC_TESS Models}}
\begin{tabular}{lcc}
\hline
Parameter   & Hot spot model & Cool spot model \\
\hline
$i$ ($\degr$)   & 51.399 $\pm0.106$ & 52.422 $\pm0.133$  \\
$T_1$ (K)       & 10000 (fixed)     & 10000 (fixed)     \\
$T_2$ (K)       & 6226 $\pm150$     & 7409 $\pm200$     \\
$\Omega_1 = \Omega_2$   & 2.1374 $\pm0.0014$    & 2.1198 $\pm0.0011$ \\
$q=M_2/M_1$     & 0.174 (fixed)     & 0.174 (fixed)     \\
Fill-out factor & 0.27              & 0.42   \\
$r_1$ (volume)  & 0.542 $\pm0.002$  & 0.549 $\pm0.002$  \\
$r_2$ (volume)  & 0.250 $\pm0.001$  & 0.256 $\pm0.001$ 	\\
$L_1/(L_1+L_2)$ & 0.93 $\pm0.03$    & 0.90 $\pm0.03$	\\
$L_2/(L_1+L_2)$ & 0.07 $\pm0.01$    & 0.10 $\pm0.01$     \\
$l_3$           & --                & 0.09 $\pm0.02$     \\
\hline
\end{tabular}
\end{center}
\end{table}

The most significant difference between the
parameter sets given in
Table \ref{Tab:WD+MC_TESS Models} is in the effective temperature of the secondary component ($T_2$) and fill-out factor ($f$). The hot spot model reduced $T_2$ to 6000 K, with $f$ at 0.25; while the cool spot increased $T_2$ to 7500 K wirh $f$ at 0.40. Naturally, physical interpretation of  the data should be influenced by the `goodness of fit' of parameter-set results appropriate to such modelling.

The $\chi^2$ test yields similar results for both models,  though the cool spot appears slightly better (see Tables \ref{Tab:WD+MC_TESS_hotspot_Model} and \ref{Tab:WD+MC_TESS_coolspot_Model}). In the cool spot fitting, two additional parameters (spot latitude and third light $l_3$)  allowed more freedom compared to the hot spot approach. 
It is well known that the more parameters included in a model, the better will be the fit.  Even so, the process of adding more parameters will conflict with the determinacy of the model when the information limit is exceeded.

Although the colour curves in the top panel of Fig.~\ref{fig:wd_coolspot_model} are not very distinct, they offer support to the cool spot model. On the other hand, the quadratic O--C variation in the bottom panel of Fig.~\ref{fig:o-c_all} indicates a decrease in the orbital period, which is associated with mass transfer between the components. Such interaction may lead to localized heating on the secondary's surface, consistent with the presence of a hot-spot.

In this way, a small scale of disequilibrium  can be accommodated within  the concept of 'poor thermal contact' that gives rise to disequilibrium effects on a Kelvin time-scale. Energy and mass transfer within the common envelope account for  surface flows and localized temperature variations.
 
Systematic  effects in the residual light curves are thus explained (see Figs.~\ref{fig:wd_hotspot_model} and \ref{fig:wd_coolspot_model}). Regarding the quasi-periodic jitter effect discussed in Section~\ref{O_Connell}, these may also be related to surface flows and interaction effects. Adherence to the static equipotential construction loses strictness, allowing intermittency  in the  mass-transfer.

In conclusion, most of the available data supports the hot spot model. However, more sensitive spectral data are needed to draw definitive conclusions.

\subsection{WF fits to TESS LCs}
\label{sec:wf_tess_fits}

A fuller perspective of the LC analyses is obtained by applying {\sc WF} to the TESS photometry, represented by a binned sample from the five sectors discussed on Section~\ref{sec:tess_obs},  presented as 1000 normalized flux values  uniformly distributed over the phase range 0 -- 1.

\begin{figure}
\centering
\includegraphics[width=\linewidth]{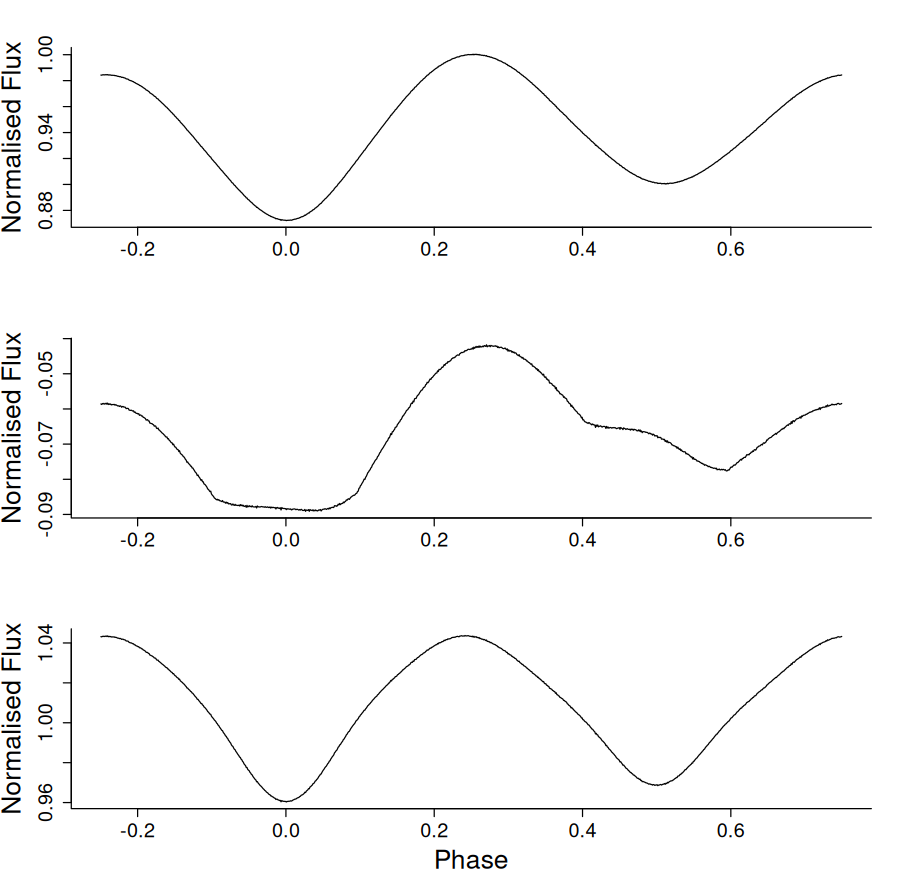}  
    \caption{Top panel: TESS LCs averaged from the 5 sectors discussed in Section \ref{sec:wd+mc fits},  with near-contact model initial fitting. The systematic residuals from the initial fit can be seen in the middle panel. Bottom panel: {\sc WF} near-contact model optimal fit to the TESS data `cleaned' of systematic residuals. } 
    \label{fig:wf_model}   
\end{figure}

It was noted in \cite{Erdem_2025} that {\sc WD} and {\sc WF} converge to the same fitting function with low structural coefficients $k_i$ and truncation of the  series expansion for the tidal deformation at $i = 4$. (the octupole approximation). The parameters in {\sc WF}, introduced in Section \ref{sec:bv_observations}, are listed again in Table~\ref{Tab:WF_TESS Model}, the fitting now being constrained by the  spectroscopic mass ratio  $q = 0.174$ (Section \ref{sec:spectroscopy}). The elimination of O'Connell effect systematic errors, shown in the middle panel of Fig.~\ref{fig:wf_model}, also constrain the fitting to apply better to the close binary system alone.  The middle panel of Table~\ref{Tab:WF_TESS Model} lists the leading terms of the `cleaning' Fourier series, together with a representative uncertainty.   

The parameter meanings in Table~\ref{Tab:WF_TESS Model} are essentially similar to those  given above in the {\sc WD} program,  Instead of the quantity $\Omega$, however, which locates a tidally  distorted surface in the {\sc WD} program, the representative `side' relative radius $r$ is used. This radius is styled $r_0$ in \cite{Kopal_1959}'s notation. It is connected to the  potential though the equation
\begin{equation}
r_0 = \frac{1}{(\Omega_K - q)}  \,\,
\end{equation} 
where $\Omega_K$ is Kopal's  slightly modified version of the formal Roche potential $\Omega$ in the rotating coordinate frame of the (presumed circular) orbit (\citeauthor{Budding_2022}, \citeyear{Budding_2022}, Ch.\ 4, Eqn 4.3). Results of the LC fitting experiments are presented in Table~\ref{Tab:WF_TESS Model} and displayed in Fig~~\ref{fig:wf_model}.

Table~\ref{Tab:WF_TESS Model} lists also primary and secondary effective temperatures $T_{e1}$ and $T_{e2}$, that are obtained by adopting the Rayleigh-Jeans approximation for the representative fluxes from either star so that 
\begin{equation}
\label{eq:R_J}
\frac{T_{e2}} { T_{e1}} = 
 \frac{L_2}{k^2 L_1}  ,
\end{equation}
where the ratio of the radii $k = r_2/r_1$. 

We should notice here a difference between these temperatures and the local equilibrium temperatures corresponding to the barotropic approximation that should hold if the stellar surfaces correspond to equipotentials in accordance with Maclaurin's principle \citep[see, e.g.,][]{Kopal_1959,Chandrasekhar_1969}.
The surface temperatures of contact binaries are thus expected, and generally found, to be similar in value, having good `thermal contact', even though the mass ratio, for measured examples, usually appreciably less than unity.  In the present case, the temperature difference is rather conspicuous. While asymmetry in the flow regime of this unusual near-contact binary may have some part to play in explaining the temperature difference, the high effect of errors in both small numerator and denominator on the right in Eq.~\ref{eq:R_J} cannot be overlooked.

The role of errors arises again in Table~\ref{Tab:WF_TESS Model} regarding the quantities Rochefill (1) and (2). These are derived using Tables 4.2 and 4.3 in \cite{Budding_2022}, which for the mass ratio $q = 0.174$ yield $r_1 = 0.5420$, and $r_2 = 0.2328$. This suggests the primary is very close to its surrounding Roche lobe, with a slightly detached secondary.  On the other hand, if the mass ratio were to have the slightly lower value of $q = 0.15$ the secondary would be in contact with its Roche lobe with the primary detached. Clearly, the assigned mass ratio is of critical importance 
for the stars' physical  arrangement.
 
\begin{table}
\begin{center}
\caption{Results of the {\sc WF} fitting to the binned TESS LC of V486 Car.
\label{Tab:WF_TESS Model}}
\begin{tabular}{ll}
\hline
Parameter                     & Value \\
\hline
$U$                           & $0.99017 \pm 0.00002$  \\
$L_1$                         & $ 0.903\pm0.03$	\\
$L_2$                         & $ 0.097\pm0.008$ \\
$r_{1}$ 	                  &  0.545$0\pm0.001$	\\
$r_{2}$	                      &  0.2214$\pm0.004$	\\
$i$ ($\degr$)                 & $51.40\pm0.15$    	\\
$\Delta \phi_0$               & $0.005\pm0.001$   	\\
$q=M_2/M_1$ 	              &  0.174 (fixed)  \\
$T_{e1}$ (K)                  &  10000 (fixed) \\
$T_{e2}$ (K)                  &  6200$\pm 140$	\\
Rochefill (1)                &  +0.0035 \\
Rochefill (2)                &  --0.0114 \\
\hline
\multicolumn{2}{c}{Fourier coefficients. for systematic residuals ($\times 10^{-4} )$} \\
$ a_0 $  &  --675 $\pm 4$ \\
$ a_1, a_2,a_3, a_4  $        &  --88,--151, --14, 30 ... \\
$ b_1, b_2, b_3, b_4 $        &    75, --33, --4, --2 ...  \\
\hline
$\Delta l$ mean datum error  	                 &		0.0015 \\
$\nu$  No. degr. freedom     	                 &	  	994	\\
$\chi^{2} / \nu $                                &	 	1.08 \\
\hline												
\end{tabular}
\end{center}
\end{table}

\subsection{WD+MC fits to B, V and Hipparcos LCs}
\label{sec:wd_bvhp_fits}

We also used the WD+MC program for fittings of the Hipparcos and our B, V LCs.  The input ranges of the adjustable parameters ( $\Delta \phi $, $i$, $T_2$, $\Omega_{1,2}$ and $L_1$) were set, as with the TESS  LCs (Section~\ref{sec:wd+mc fits}).  The large O'Connell effect is noticeable in both the BV LCs and the Hipparcos LCs, appearing at almost the same phases as in the TESS LCs. The spot parameters were thus added to the WD+MC program, and their input ranges were set the same as with the TESS LCs.

Since the scatter in ground-based B and V observations, the WD+MC program gave the geometric elements quite different from the results of TESS LC analysis (Table~\ref{Tab:WD+MC_TESS Models}). Therefore, the geometric parameters ($i$, $\Omega_{1,2}$) were fixed at those found in the TESS LC analyses and just $T_2$, phase shift, $L_1$ and spot parameters were left free.

\begin{figure}
\centering
    \includegraphics[scale=0.55]{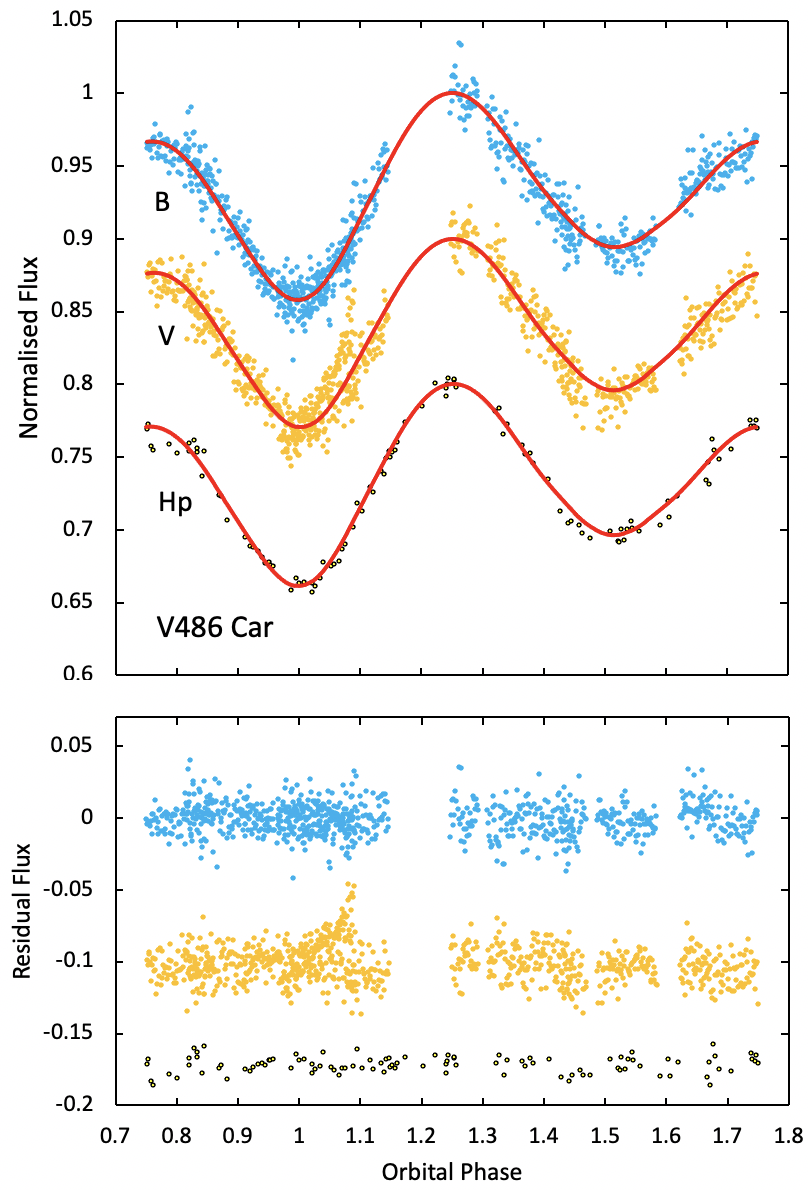}
    \caption{B, V and Hipparcos (Hp) LCs with the WD+MC model fitting. Residuals to the LC model are plotted in the lower figure. } 
    \label{fig:wd_model_bvhp}
\end{figure}

The adopted parameters of the WD+MC model for the B, V and Hipparcos LCs are presented in Table~\ref{Tab:WD+MC BV&Hp Model}. A comparison of WD+MC model fitting with B, V and Hipparcos LCs is shown in Figure~\ref{fig:wd_model_bvhp}. 

\begin{table}
\caption{Results of the WD+MC fitting to the B, V and Hipparcos (Hp) light curves of V486 Car.
\label{Tab:WD+MC BV&Hp Model}}
\scalebox{0.90}{
\begin{tabular}{lccc}
\hline
Parameter   & B & V & Hp \\
\hline
$\Delta \phi$           & $0.0077\pm0.0007$     & $0.0100\pm0.0001$     & $0.0100\pm0.0013$ \\
$i$ ($\degr$)           & 51.399 (fixed)        & 51.399 (fixed)        & 51.399 (fixed)	\\
$T_1$ (K)               & 10000 (fixed)         & 10000 (fixed)         & 10000 (fixed)   \\
$T_2$ (K)               & 5700$\pm159$ 	        & 6055$\pm126$          & 5720$\pm346$ \\
$\Omega_1 = \Omega_2$   & 2.1374 (fixed) 	    & 2.1374 (fixed)        & 2.1374 (fixed)	\\
$q=M_2/M_1$ 	        & 0.174 (fixed)         & 0.174 (fixed)         & 0.174 (fixed)   \\
Fill-out factor         & 0.26                  & 0.26                  & 0.26 \\
$r_1$ (volume)	        & 0.542 (fixed)         & 0.542 (fixed)         & 0.542 (fixed)	\\
$r_2$ (volume)	        & 0.250 (fixed)         & 0.250 (fixed)	        & 0.250 (fixed)	\\
$L_1/(L_1+L_2)$         & 0.979$\pm0.033$ 	    & 0.961$\pm0.035$       & 0.969$\pm0.033$ \\
$L_2/(L_1+L_2)$         & 0.021$\pm0.005$       & 0.039$\pm0.007$       & 0.031$\pm0.005$  \\
\hline
Spot parameters \\
$\beta$ ($\degr$)       & 90 (fixed)            & 90 (fixed)            & 90 (fixed)    \\
$\lambda$ ($\degr$)     & 128 $\pm4$            & 147 $\pm8$            & 132 $\pm6$ \\
$\gamma$ ($\degr$)      & 60 $\pm4$             & 60 $\pm6$             & 60 $\pm5$ \\
$\kappa$                & 1.372 $\pm0.035$      & 1.317 $\pm0.049$      & 1.343 $\pm0.043$ \\
\hline
$\chi^{2}_{\rm red}$    &	1.24	            &	1.70	            &	1.73	\\
$\nu$      	            &	848	                &	848	                &	83	\\
$\Delta l$ 	            &	0.010	            &	0.010	            &	0.005	\\
\hline												
\end{tabular}}
\end{table}

We calculated the relative light contributions of each component to the total light of the system (as $l_1+l_2$) in each band at 0.25 phase from the results in Table \ref{Tab:WD+MC BV&Hp Model}. These $l_1$ and $l_2$ values, and their corresponding magnitudes and colours, are given in Table \ref{tab:WD_colors}. In this calculation, the magnitude $V = 6.263$ and $B$ = 6.307 mag at 0.25 phase were used. 
The unreddened colour indices were obtained from the colour excess $E(B-V) = 0.054$ mag given in Section \ref{sec:bv_observations}. 

From Table~\ref{tab:WD_colors} we see that the primary star has (B -- V)$_0$ of $-0.02$ mag, in agreement with the assigned A0V spectral type (Section~\ref{sec:bv_observations}). The (B -- V)$_0$ = 0.67 mag derived for the secondary component corresponds to a temperature of 5700 K from the calibration of \cite{Eker_etal_2018}. This temperature value is in agreement with the $T_2$ values obtained with the B-band and Hp-band LC solutions, although the V-band LC solution suggested a hotter second component of approximately 6000 K (see Table \ref{Tab:WD+MC BV&Hp Model}). On the other hand, SIMBAD gives a slightly less brighter combination V magnitude for V486 Car as 6.32. 

\begin{table}
    \caption{{Magnitudes and colours of V486 Car from WD+MC results. Errors are on the order of 0.02 mag. } }
    \centering
    \begin{tabular}{lcc}
    \hline
Parameter           & Primary           & Secondary  \\
\hline
$l_{1,2}$ (B,V)     & 0.979, 0.961      & 0.021, 0.039 \\
$B$                 & 6.333             & 10.504  \\
$V$                 & 6.303             & 9.782  \\
$B-V$               & 0.030             & 0.722  \\
$(B-V)_0$           & $-0.024$          & 0.668  \\
$V_0$               & 6.136             & 9.615 \\
\hline
    \end{tabular}
\label{tab:WD_colors}
\end{table}

\section{Analysis of Spectroscopic Data}
\label{sec:spectroscopy}

\subsection{Preparation of Spectral Data}

Spectroscopic data on V486 Car were gathered over 13 nights during the period December 2010 to December 2014 (see Table~\ref{tab:observing_log}) with the High Efficiency and Resolution Canterbury University Large \'{E}chelle Spectrograph (HERCULES) of the Department of Physics and Astronomy, University of Canterbury  \citep{Hearnshaw_2002}. This was attached to the 1m McLellan telescope at the Mt John Observatory (UCMJO).  Images were collected with a 4k$\times$4k Spectral Instruments (SITe) camera \citep{Skuljan_2004}. The 100 $\mu$m fibre, which is suited to typical seeing conditions at Mt John, enables a theoretical resolution of $\sim$40,000. Raw observations were reduced with the latest version of the software {\sc hrsp} \citep{Skuljan_2021}, that produces wavelength calibrated and normalized output conveniently in `FITS'\footnote{See https://fits.gsfc.nasa.gov/standard40/fits\_standard40aa-le.pdf for further details on this file format.} formatted files.  Each recording covers the wavelength range 4520-6810 {\AA}.

The HERCULES spectra exhibit order-to-order variations in the original dispersion, ranging from 0.02~\AA/pixel in the bluest order (127) to 0.039~\AA/pixel in the reddest order (79).  For the analysis presented here, we selected spectral orders based on the line densities listed in Table~\ref{tab:spectral_features}: orders 92, 107, 109, 110, 113, 124, and 125.  To mitigate edge effects, the outer regions of each order were excluded, and within each order regions containing prominent metallic lines were specifically cropped.  The cropped spectral regions were resampled to contain 1024 data points, resulting in the following velocity-dispersion sampling: 1.56~km\,s$^{-1}$ (order~92), 2.32~km\,s$^{-1}$ (order~107), 2.04~km\,s$^{-1}$ (order~109), 2.45~km\,s$^{-1}$ (order~110), 1.80~km\,s$^{-1}$ (order~113), 2.73~km\,s$^{-1}$ (order~124), and 3.16~km\,s$^{-1}$ (order~125).  A total of 25 spectra per order, as summarized in Table~\ref{tab:observing_log}, were subsequently analyzed using \textsc{korel} \citep{Hadrava1995}.

\subsection{Determination of the Optimal Radial-Velocity Semi-Amplitudes}

To determine the optimal radial-velocity semi-amplitudes of the system, we carried out a two-dimensional parameter scan over the ranges  $K_{1} = 15$--$45~\mathrm{km\,s^{-1}}$ and  $K_{2} = 60$--$220~\mathrm{km\,s^{-1}}$,  each sampled with ten uniformly spaced grid points.   For every $(K_{1}, K_{2})$ pair, the corresponding entries in the \texttt{korel.par} file were updated, setting $K_{1}$ directly and computing the mass ratio $q = K_{1}/K_{2}$. The \textsc{korel} disentangling algorithm was then carried out. For each run, two independent solution-quality indicators were extracted, i.e.;

\begin{enumerate}
    \item \textbf{Residual root mean square (RMS) from \texttt{korel.res}}:  
    The internal RMS reported by \textsc{korel}, reflecting the global quality of the disentangling solution.

    \item \textbf{Block-wise sigma values from \texttt{korel.o-c}}:  
    Each O--C (i.e.\ residual) block was read and its standard deviation computed.  
    The mean of all block sigmas was adopted as an external, observationally driven RMS estimator.
\end{enumerate}

\begin{figure}
        \begin{center} 
          \includegraphics[width=1.0\columnwidth]
          {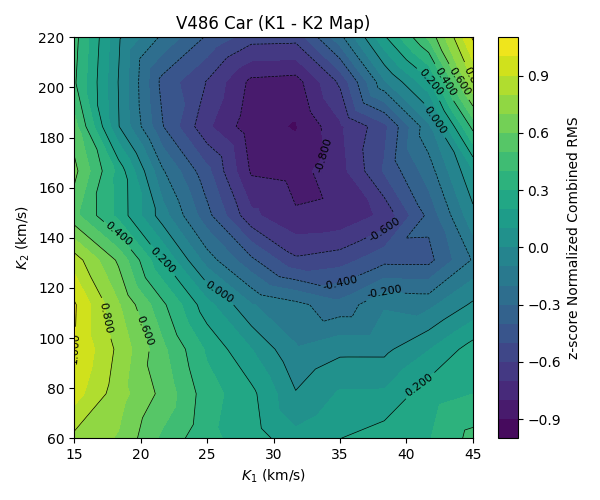}
        \end{center}
\caption{Two–dimensional z-score–normalized RMS map for the $K_1–K_2$ parameter grid. Each point represents the combined RMS value obtained from both the O–C residuals and the \textsc{korel} correlation output. The RMS values are independently z-score normalized and averaged to produce a unified goodness-of-fit metric across the grid. Darker regions indicate parameter combinations yielding lower normalized RMS, thus representing better overall solutions. The global minimum marks the optimal pair ($K_1$, $K_2$) adopted in this study.}
\label{fig:chi_map}
\end{figure}

Because these two metrics have different numerical scales, each RMS set was normalized using a z-score transformation. Here, 'z-score normalization' (sometimes informally called z-scaling) means subtracting the sample mean and dividing by the standard deviation, so that each metric is placed on a common scale with mean 0 and variance 1. This step ensures that neither of the two RMS indicators dominates the combined quality metric due to its absolute magnitude. The final quality metric for each grid point was defined as the average of the two normalized values.   The resulting $K_{1}$--$K_{2}$ map, shown in Fig~\ref{fig:chi_map} illustrates the combined normalized RMS surface and allows the global minimum to be identified. This minimum is adopted as the best-fitting pair of semi-amplitudes, while the surrounding contour levels provide a natural basis for estimating the parameter uncertainties.

\begin{figure}
        \begin{center} 
          \includegraphics[width=1.0\columnwidth]{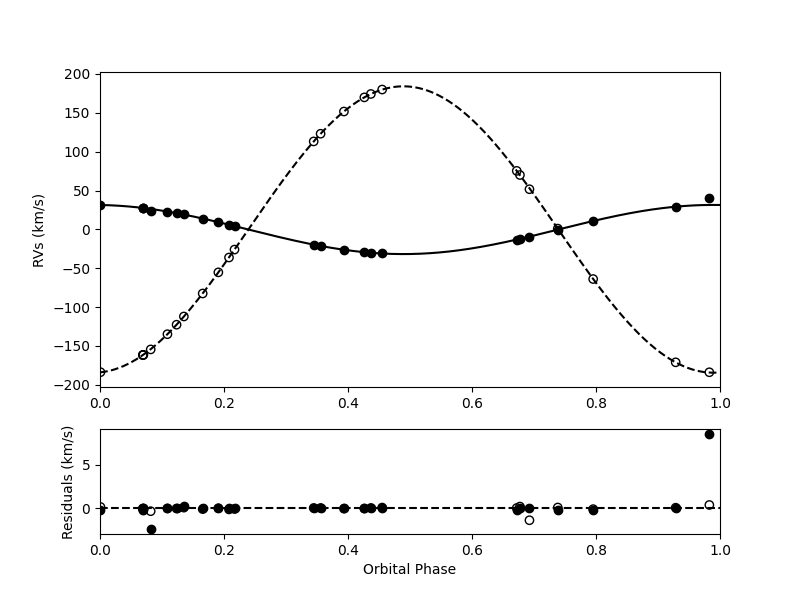}
        \end{center}
\caption{RVs and the adopted spectroscopic orbit.}
\label{fig:RV_orbit}
\end{figure}

In this analysis we did not use the regression-style RMSE metric, as the KOREL disentangling procedure does not involve a regression fit in the usual sense, but instead evaluates the consistency between the observed and reconstructed spectra through residuals \citep{Hadrava1995,Hadrava2004}. Instead, we evaluate the solution quality using two physically motivated RMS indicators: (1) the internal residual RMS reported by \textsc{korel}, and (2) the block-by-block standard deviations computed from the observed minus computed (O–C) spectra. These two RMS measures quantify different aspects of the spectral reconstruction rather than statistical regression error. Because they naturally have different numerical scales, each RMS distribution was standardized using a z-score normalization (mean subtraction and division by the standard deviation) before being combined into a single quality metric. The orbital solution, the orbital parameters, and the measured RVs are presented in Fig.\ref{fig:RV_orbit}, Table~\ref{tab:orbit_pars}, and Table~\ref{tab:RVs}, respectively.

\begin{table}
\begin{center}
\caption{Parameters of the spectroscopic orbit.
\label{tab:orbit_pars}}
\footnotesize
\begin{tabular}{lcl}
  \hline 
\multicolumn{1}{l}{Parameter}  & \multicolumn{1}{c}{Value} &
\multicolumn{1}{l}{Error}  \\
\hline
\multicolumn{1}{l}{Period (days)}  & \multicolumn{1}{c}{1.0938842} &
\multicolumn{1}{l}{--}  \\
\multicolumn{1}{l}{$T_0$ (HJD)}  & \multicolumn{1}{c}{2460747.7636} &
\multicolumn{1}{l}{0.0070}  \\
\multicolumn{1}{l}{$K_1$ (km/s)}  & \multicolumn{1}{c}{32.1} &
\multicolumn{1}{l}{0.4}  \\
\multicolumn{1}{l}{$K_2$ (km/s)}  & \multicolumn{1}{c}{184.2} &
\multicolumn{1}{l}{0.4}  \\
\multicolumn{1}{l}{e}  & \multicolumn{1}{c}{0.0} &
\multicolumn{1}{l}{--}  \\
\hline 
\end{tabular}
\end{center}
\end{table}

\begin{table}
\begin{center}
\caption{Measured RVs and errors.
\label{tab:RVs}}
\footnotesize
\begin{tabular}{ccrrrr}
  \hline 
\multicolumn{1}{l}{HJD}  & \multicolumn{1}{c}{Phase} &
\multicolumn{1}{l}{RV1} & \multicolumn{1}{l}{errRV1} & \multicolumn{1}{l}{RV2} & \multicolumn{1}{l}{errRV2}\\
\multicolumn{1}{l}{}  & \multicolumn{1}{c}{} &
\multicolumn{1}{l}{(km/s)} & \multicolumn{1}{l}{(km/s)} & \multicolumn{1}{l}{(km/s)} & \multicolumn{1}{l}{(km/s)}\\
\hline
2455542.05680       &0.081      &27.4       &0.6        &$-161.5$    &3.5 \\
2455542.06970       &0.092      &23.9       &5.8        &$-154.5$   &4.1 \\
2455544.90330       &0.683      &$-13.3$    &1.2        &75.3       &6.8 \\
2455544.92570       &0.703      &$-9.2$     &1.2        &51.7       &6.8 \\
2455544.97570       &0.749      &$-0.5$     &1.3        &1.0        &7.5 \\
2455545.03810       &0.806      &10.6       &1.3        &$-63.9$    &7.7 \\
2455795.83700       &0.080      &27.6       &0.6        &$-161.8$   &3.5 \\
2455795.88040       &0.119      &23.0       &0.9        &$-134.9$   &5.0 \\
2455795.90940       &0.146      &19.4       &1.2        &$-112.1$   &5.9 \\
2455795.94260       &0.176      &14.0       &1.1        &$-82.5$    &6.6 \\
2455795.98900       &0.219      &6.1        &1.3        &$-36.2$    &7.3 \\
2455796.19180       &0.404      &$-26.2$    &0.7        &151.7      &4.3 \\
2455796.22780       &0.437      &$-29.3$    &0.5        &169.9      &3.0 \\
2455798.02490       &0.080      &27.6       &0.6        &$-161.7$   &3.5 \\
2455798.08430       &0.134      &20.9       &0.9        &$-122.7$   &5.5 \\
2455798.15800       &0.202      &9.4        &1.2        &$-55.3$    &7.1 \\
2455798.18650       &0.228      &4.3        &1.3        &$-26.0$    &7.4 \\
2455875.99200       &0.355      &$-19.5$    &1.0        &113.2      &5.9 \\
2455876.00420       &0.366      &$-21.2$    &1.0        &123.1      &5.6 \\
2455876.09290       &0.448      &$-30.0$    &0.4        &174.3      &2.5 \\
2455876.11280       &0.466      &$-31.0$    &0.3        &180.0      &1.7 \\
2455882.10010       &0.939      &29.3       &0.5        &$-171.1$   &2.9 \\
2455882.15950       &0.993      &40.0       &22.3       &$-183.9$   &0.8 \\
2455882.17910       &0.011      &31.2       &0.6        &$-183.8$   &0.4 \\
2455882.91920       &0.688      &$-12.1$    &1.2        &70.0       &6.9 \\
\hline 
\end{tabular}
\end{center}
\end{table}

\begin{figure*}
    \centering
    
    \begin{subfigure}{0.46\textwidth}
        \centering
        \includegraphics[width=\textwidth]{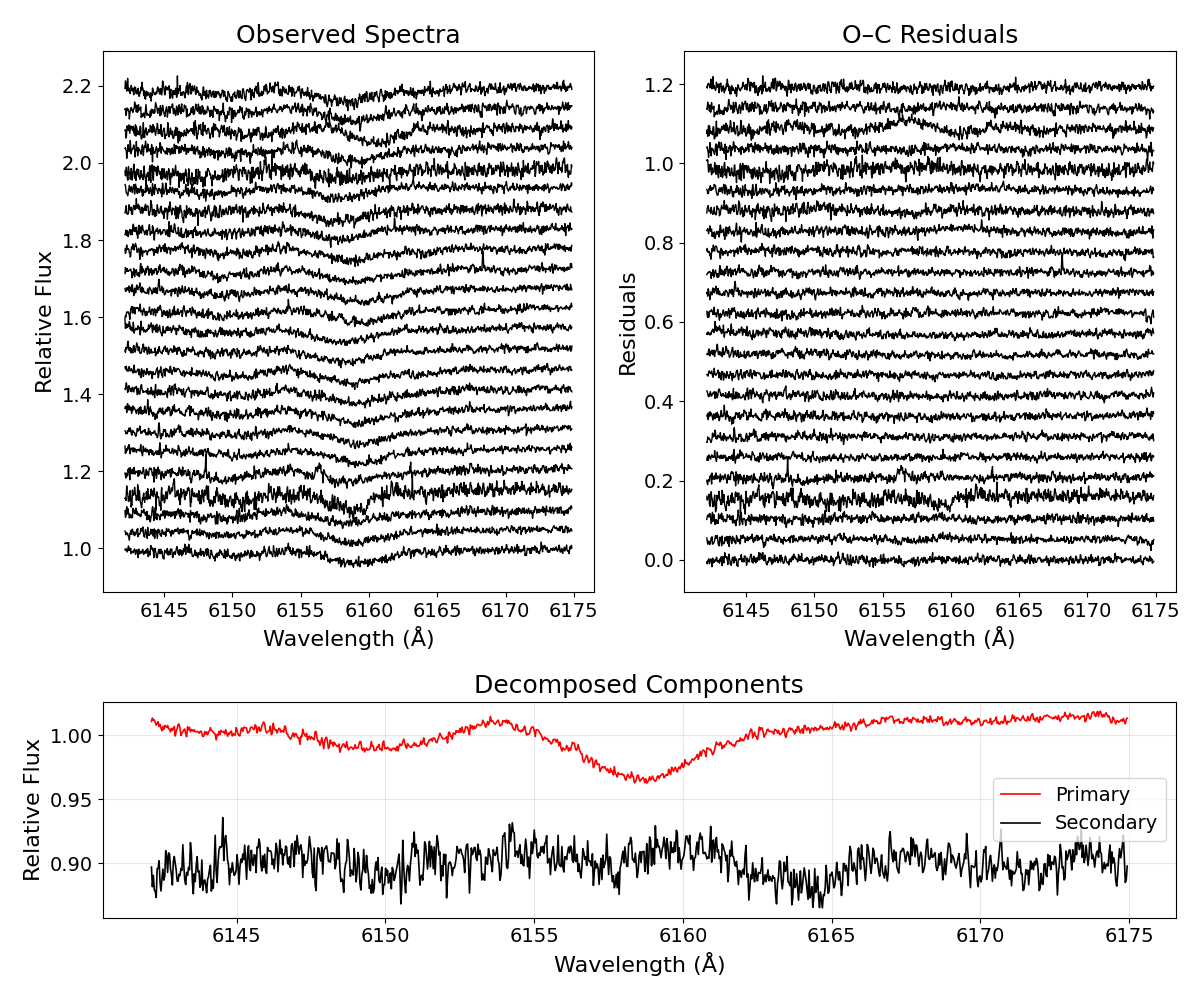}
        \caption{Order 92.}
    \end{subfigure}
    \hfill
    \begin{subfigure}{0.46\textwidth}
        \centering
       \includegraphics[width=\textwidth]{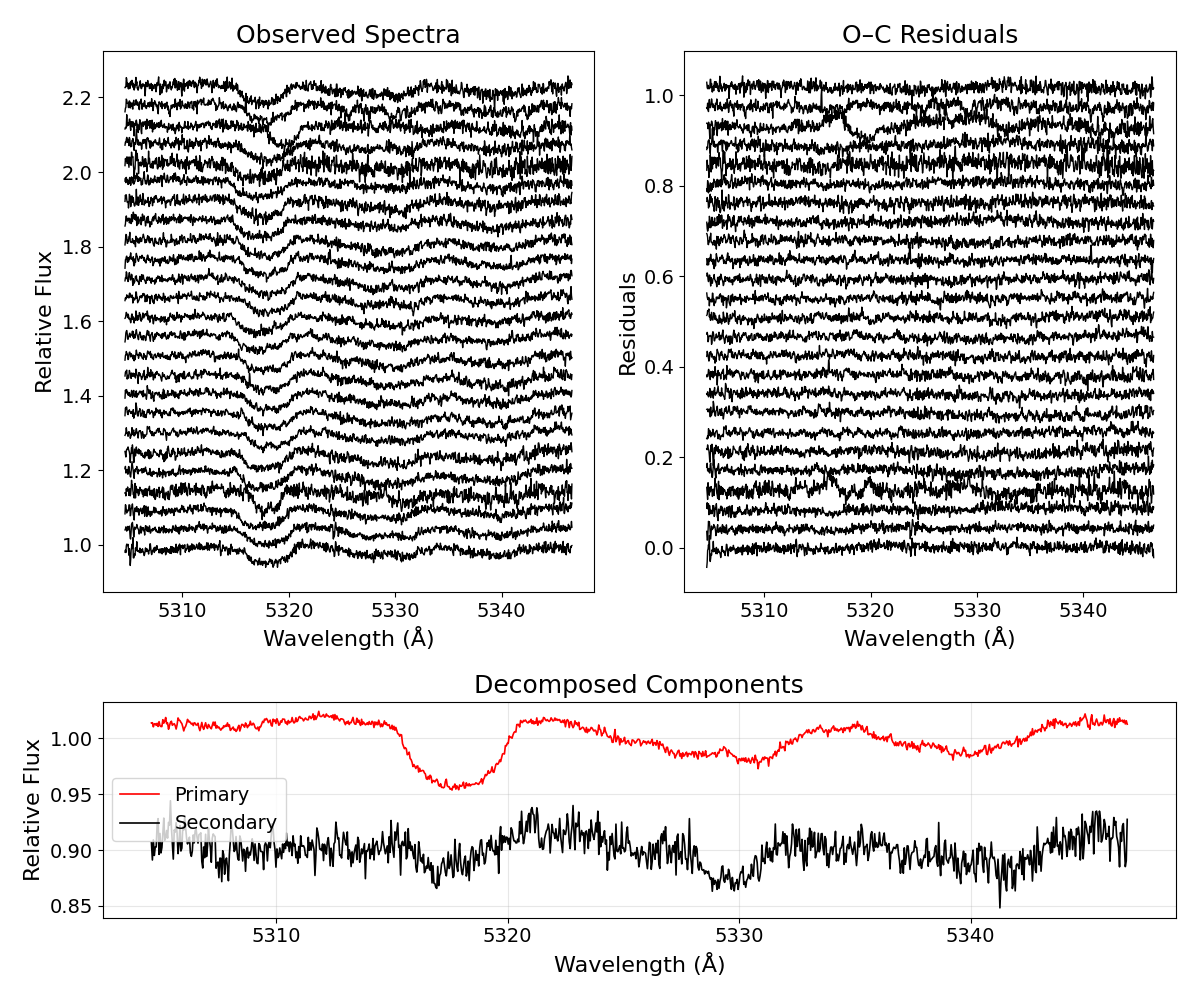}
        \caption{Order 107.}
    \end{subfigure}


    \begin{subfigure}{0.46\textwidth}
        \centering
        \includegraphics[width=\textwidth]{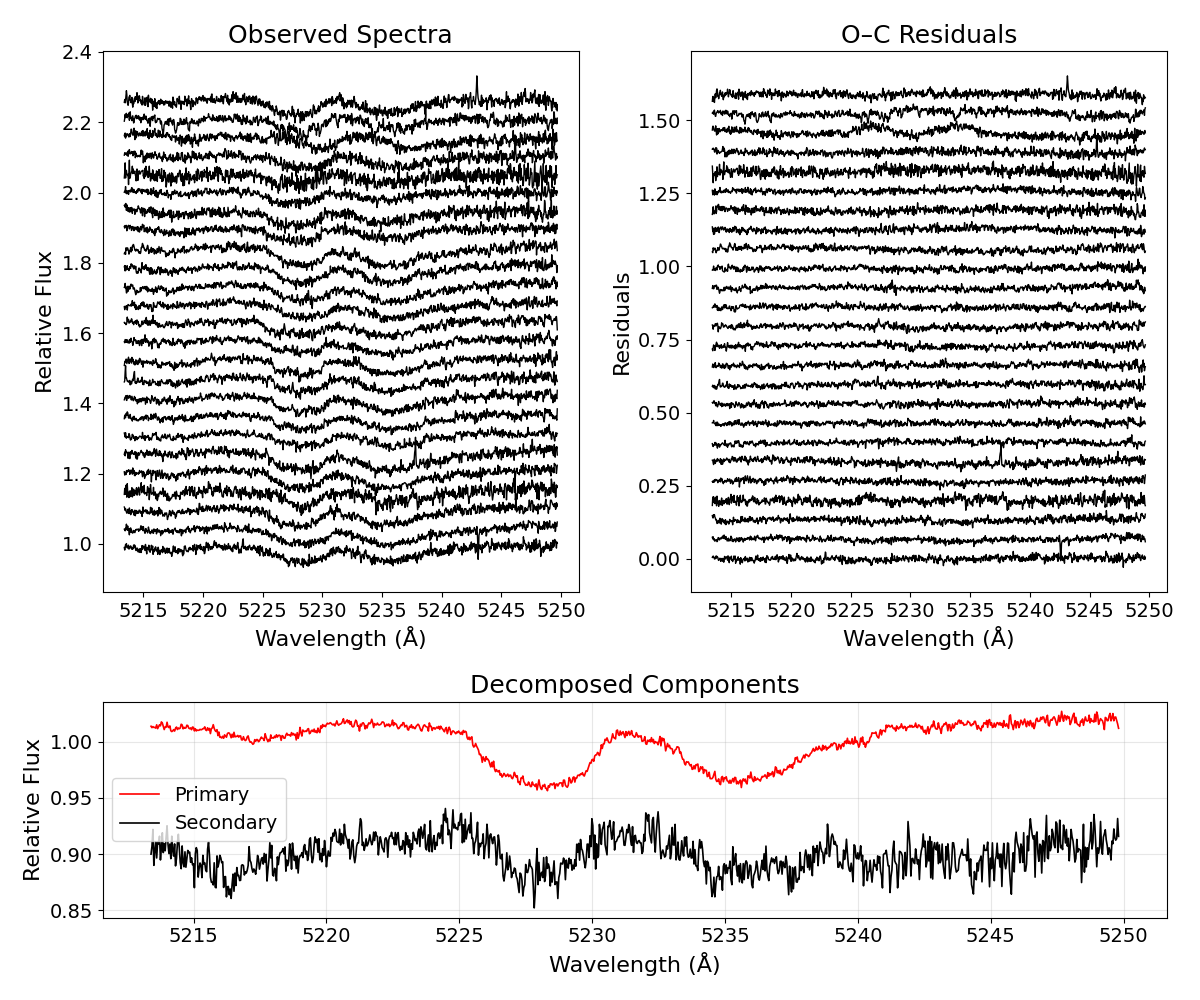}
        \caption{Order 109.}
    \end{subfigure}
    \hfill
    \begin{subfigure}{0.46\textwidth}
        \centering
        \includegraphics[width=\textwidth]{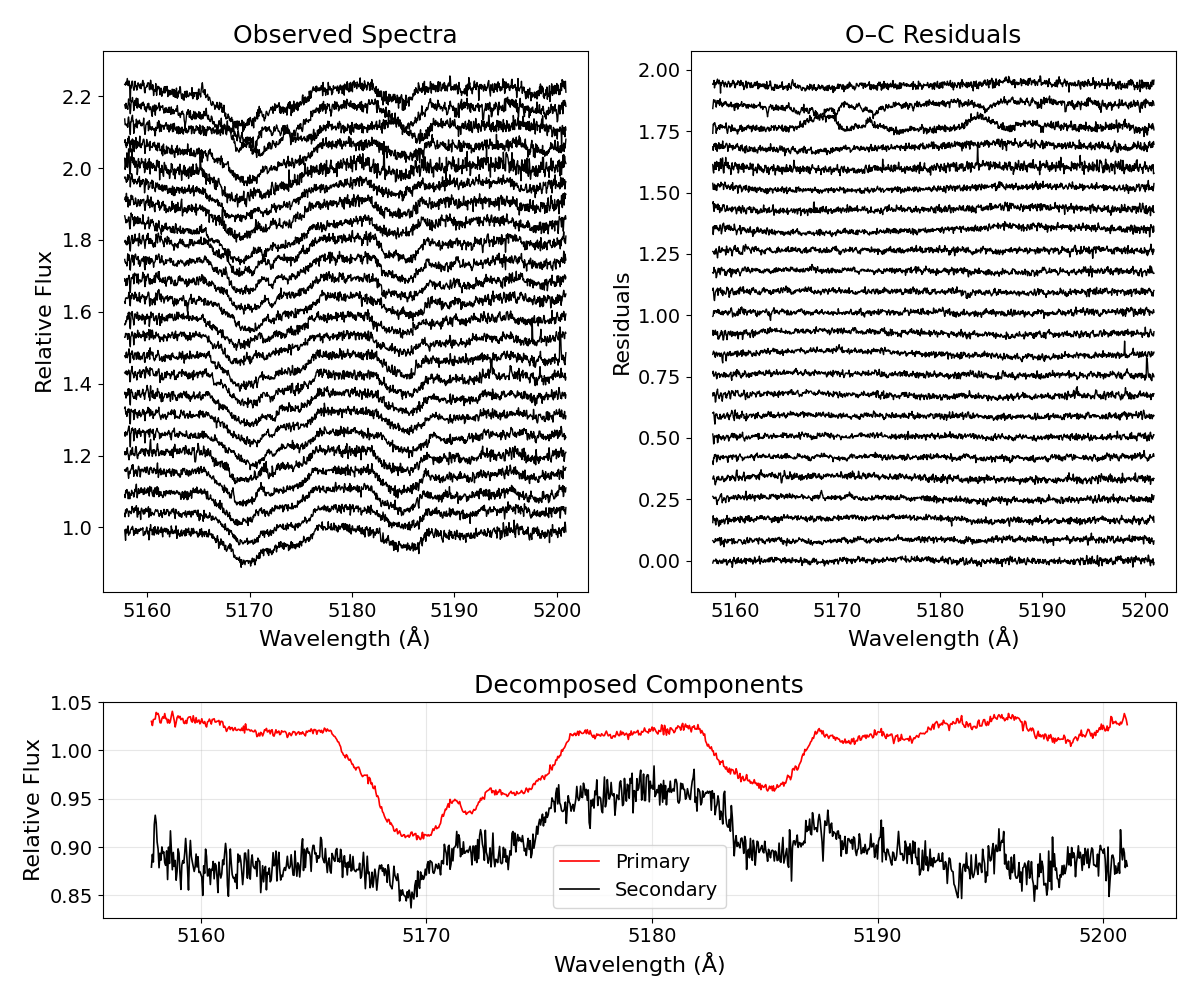}
        \caption{Order 110.}
    \end{subfigure}


    \begin{subfigure}{0.46\textwidth}
        \centering
        \includegraphics[width=\textwidth]{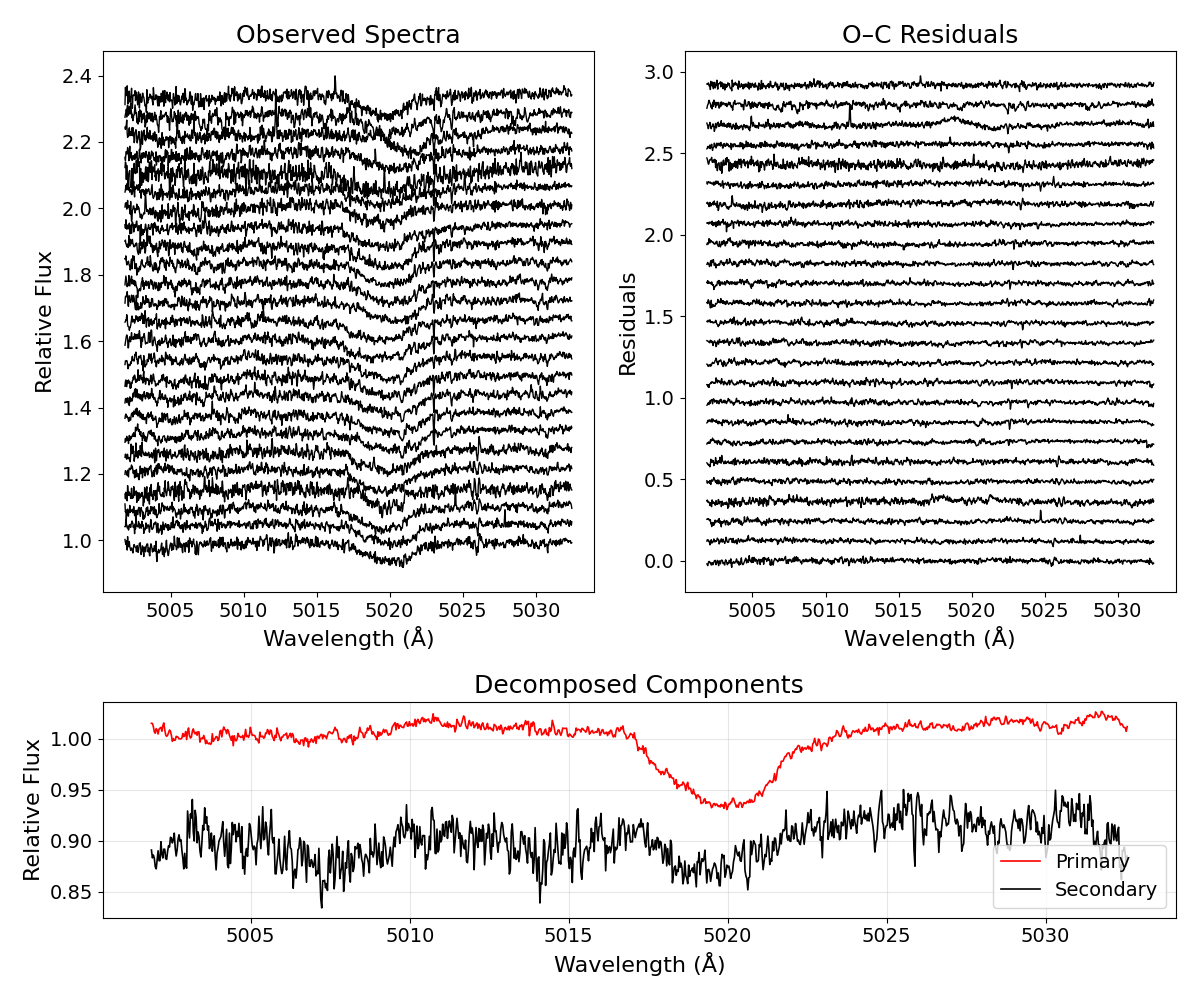}
        \caption{Order 113.}
    \end{subfigure}
    \hfill
    \begin{subfigure}{0.46\textwidth}
        \centering
        \includegraphics[width=\textwidth]{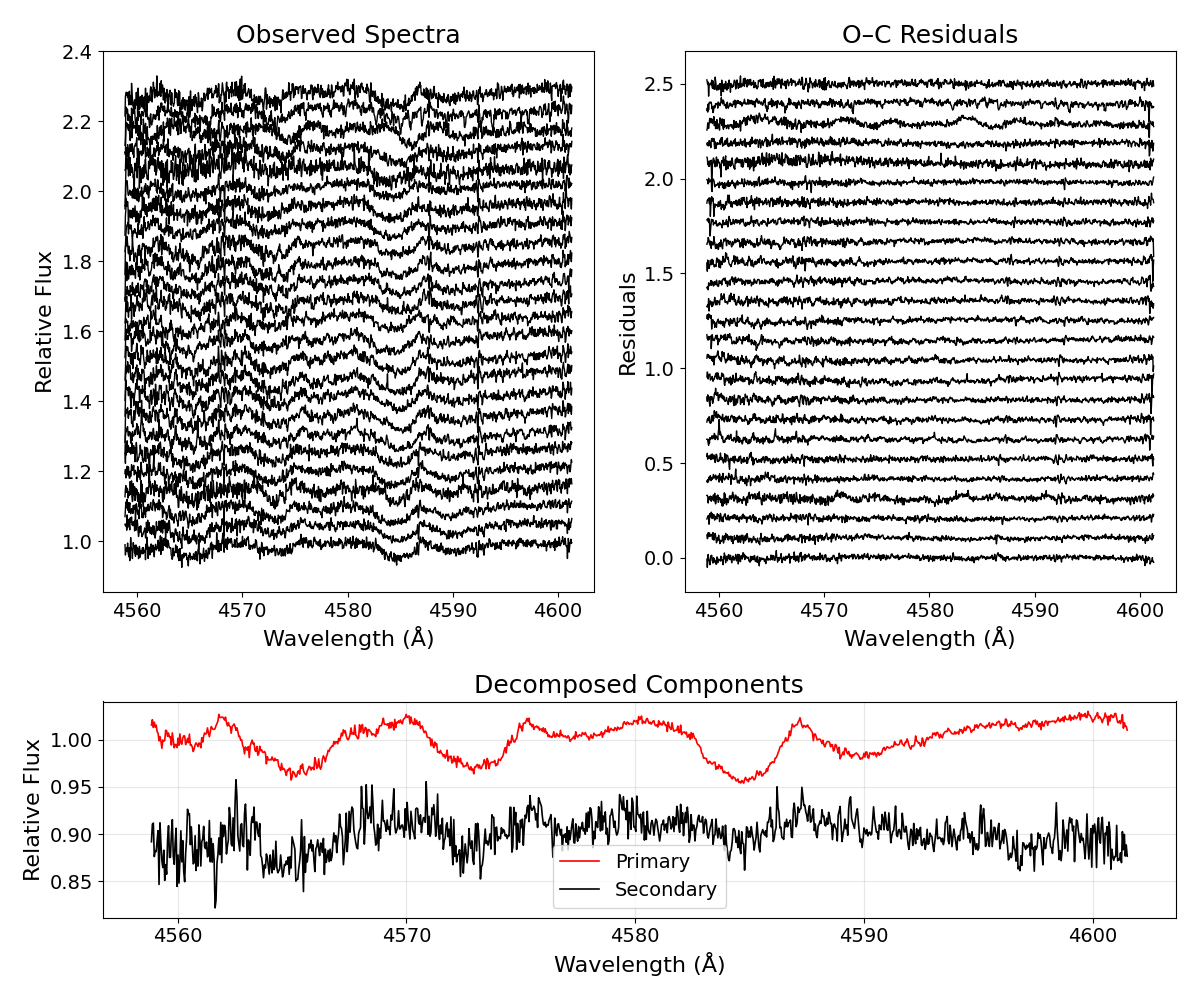}
        \caption{Order 124.}
    \end{subfigure}

\caption{Observed spectra, corresponding residuals from the \textsc{korel} analysis, and decomposed components for six datasets. In the top row, the left panel shows the observed flux as a function of wavelength, while the right panel displays the residuals of the \textsc{korel} fit. The bottom panel presents the decomposed primary and secondary components for each dataset, illustrating the contribution of individual components and the quality of the fit.}
\label{fig:residuals_comps}
\end{figure*}

\begin{figure}
        \begin{center} 
          \includegraphics[width=1.0\columnwidth]{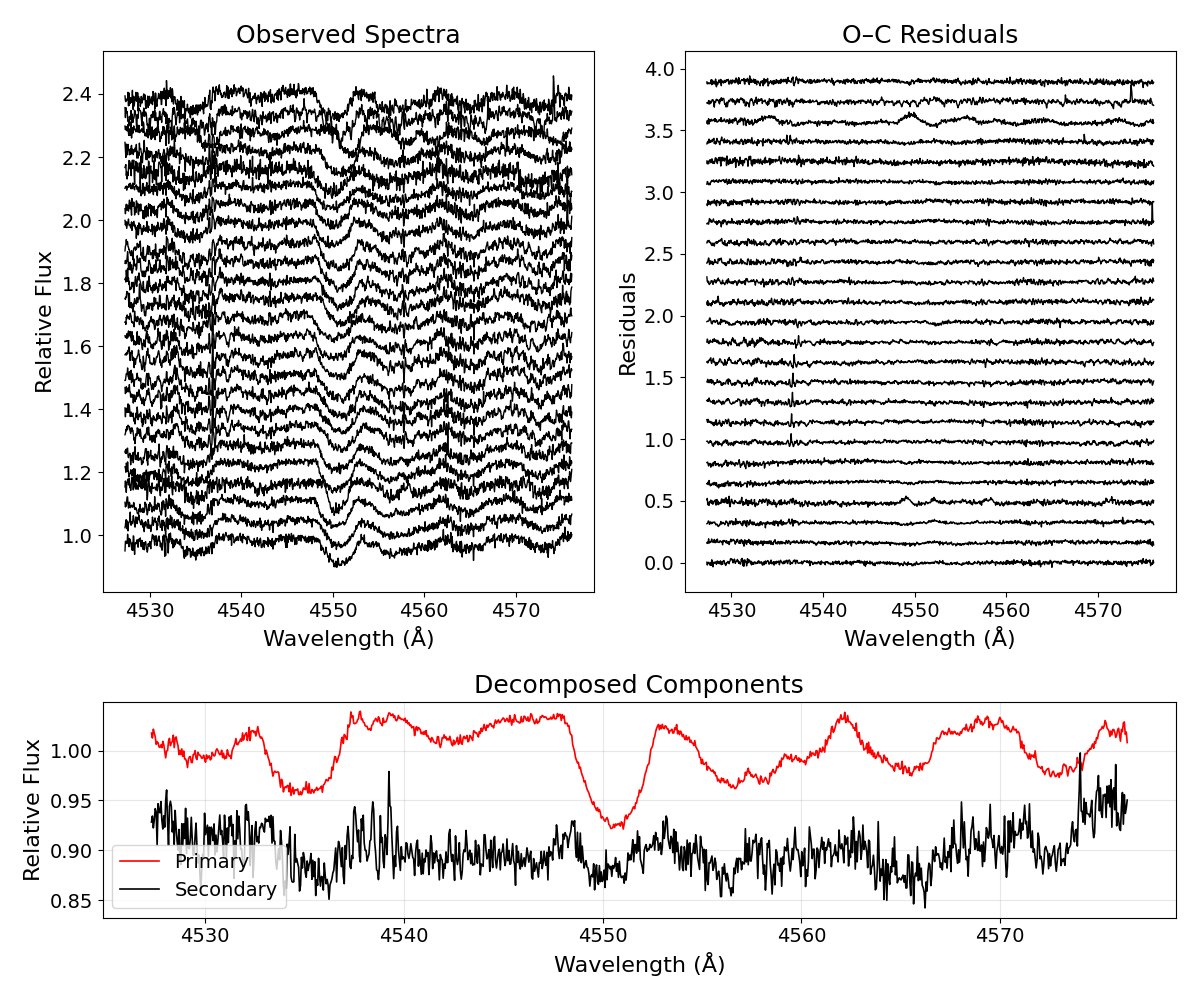}
        \end{center}
\caption{Same as Fig.\ref{fig:residuals_comps} but for spectral order 125.}
\label{fig:o125}
\end{figure}

The program {\sc Prof} \citep[latest version,][]{Erdem_2022} was used to estimate the rotational velocity of the primary component using disentangled spectral lines. {\sc Prof}, by fitting spectral lines that have sufficiently high resolution, reveals both the determination of the star's rotational velocity and information on the scale of turbulence of the plasma in which the line is formed. {\sc Prof} was applied to the disentangled metallic lines of the primary component in the orders shown in Fig.\ref{fig:residuals_comps}.  Unfortunately, the very  low signal to noise ratio of  the secondary lines prevented a meaningful application  of {\sc Prof} to them. 

\begin{figure}
\centering
	\includegraphics[scale=0.48]{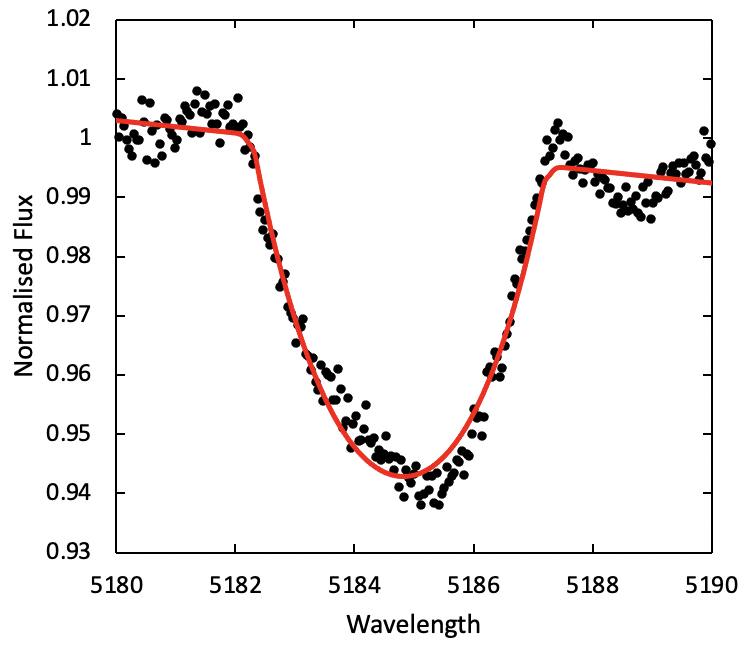}
\caption{Convolved rotation Gaussian fitting to the disentangled Mg I $\lambda$5184 line profile of the primary component of V486 Car.} 
\label{fig:prof}
\end{figure}

An example of such profile fitting is displayed in Fig. \ref{fig:prof}.
As a result, the projected mean equatorial rotation speed of the primary component was calculated as 145 $ \pm5 $ km s$^{-1}$ from the {\sc Prof} output. The turbulence velocity for the surface of the primary component was given as $\sim$2 km s$^{-1}$. 

 However, writing $v_{rot}\sin i_{rot}$ = $(2\pi R_{1}\sin i_{rot})/P_{rot}$, where we assume that the inclination of the primary star's rotation axis is equal to that of the system's orbit ($i_{rot}=i_{orb}$) and the primary has synchronised its angular rotation with the mean orbital revolution ($P_{rot}=P_{orb}$, we find the projected rotation speed of the primary to be  { 116 $\pm10$ km s$^{-1}$ } using the values of $i$ and $R_1$ in Table \ref{table:abs_par}. 

There are various reasons to account  for the discrepancy  in these derived rotation speeds including: 
(i) the large and massive primary star has not yet achieved synchronised rotation; and 
(ii) significant asymmetries are noticeable in the Mg I $\lambda$5184 and other metallic spectral line profiles of V486 Car ((Figs. \ref{fig:o125} and \ref{fig:prof}). This line profile asymmetry could be caused by the relatively large tidal distortion of the star; or (iii) complex gaseous streaming and accretion structures in this near- contact system \citep[e.g.][for $\epsilon$ 
CrA]{Rucinski_2020}.  
It should be noted that the program {\sc Prof} builds in some simplifying approximations -- such as uniform rigid-body rotation of spherical star having a linear limb darkening law. 
For the pear-shaped distortion of the primary, from Table 3.4 in \cite{Kopal_1959}  or Table 4.2 in \cite{Budding_2022} the projected radius from the orbital plane is some 23\% bigger than the mean radius at elongation phase. This compares well with the 21\% measured increase of the rotation parameter of the primary.  In this situation, we urge more and better spectroscopic observations, so as to clarify the accuracy of parametrization and gain a broader understanding of the underlying physics.

\section{Absolute Parameters}
\label{sec:absolute_parameters}

The physical parameters of the components of V486 Car were calculated based on the modelling of the RVs (Table~\ref{tab:orbit_pars}) and LCs (Table~\ref{Tab:WD+MC_TESS Models} hot spot model). In these calculations, TESS LC solutions were generally preferred, given the relatively high precision of TESS data.

The total masses of the components were calculated from the following well-known rearrangement of Kepler's third law for a circular orbit:
\begin{equation}
\label{kepler}
    (M_1 + M_2) \sin^3 i = 1.03615 \times 10^{-7} P (K_1 + K_2) ^3 ,
\end{equation}
where the masses $M_1$ and $M_2$ are in solar units, the orbital period $P$ is in days, and the RV amplitudes $K_1$ and $K_2$ are in km s$^{-1}$. 

Adopting the weighted average value of orbital inclination from Table~\ref{Tab:WD+MC_TESS Models} and taking the RV amplitudes $K_1$ and $K_2$ from Table \ref{tab:orbit_pars}, we find the component masses of V486 Car to be $M_1 = 2.06 \pm 0.11$ and $M_2 = 0.35 \pm 0.04$ in solar units.

With the total mass of V486 Car at $2.41 \pm 0.15$ M$_{\odot}$ and the period 1.0938841 d, Kepler's third law yields the semi-major axis of the relative orbit as $5.91 \pm 0.01$ R$_{\odot}$. Using the weighted average value of fractional radii of the components $r_{1}=0.542$ and $r_{2}=0.250$ 
from Table~\ref{Tab:WD+MC_TESS Models}, we find $R_1 = 3.20 \pm 0.02$ and $R_2 = 1.48 \pm 0.01$. 
The surface gravitational accelerations ($g_1$, $g_2$) are then computed from the formula $g/g_{\odot}= (M/M_{\odot}) / (R/R_{\odot})^2$.  

The bolometric magnitudes ($M_{\rm bol,1,2}$) and luminosities ($L_{1,2}$) of the component stars are calculated using the following two equations derived from Pogson's formula and the absolute radii and effective temperatures listed in Table \ref{table:abs_par}: $M_{\rm bol}=M_{\rm bol,\odot}+10\log T_{\odot} - 10\log T - 5\log (R/R_{\odot})$, and $L/L_{\odot}=10^{0.4(M_{\rm bol,\odot}-M_{\rm bol})}$. The nominal solar values, which were adopted by IAU 2015 Resolutions B2 and B3, were used in these calculations. 

The TESS-band absolute magnitudes $M_{\rm TESS,1,2}$ are derived from the bolometric correction formula: $M_{\rm TESS,1,2}= M_{\rm bol,1,2} - BC_{1,2}$. Bolometric corrections for the components were taken from the study of \citet{2023MNRAS.523.2440E}, according to their effective temperatures.
The TESS-band absolute magnitude of the system is also derived using the equation:
\begin{equation}
    M_{\rm TESS,~system}=M_{\rm TESS,2} -2.5 \log \left(1+10^{-0.4\left(M_{\rm TESS,~1}-M_{\rm TESS,~2}\right)}\right).
\end{equation}
The distance to V486 Car is calculated from the distance modulus ($d = 10^{TESS - M_{\rm TESS} + 5 - A_{\rm TESS}}$). Here $TESS$ is the apparent magnitude and $A_{\rm TESS}$ is the interstellar extinction in the TESS-band.  The TESS-band apparent magnitude, $TESS$, was taken from the MAST Portal, while the extinction, $A_{\rm TESS}$, was calculated using the relation $A_{\rm TESS}=1.940~E(B-V)$, given by \citet{2023MNRAS.523.2440E}. Here, the colour excess $E(B-V)$ was taken from the calculation in Section \ref{sec:bv_observations}.

The distance to V486 Car -- including the correction for interstellar absorption -- was computed as $162 \pm$12 pc, which match the distance of $161 \pm$2 pc given by numerical-DR3 \citep{Gaia_2022}, thus supporting the reliability of our absolute parameter results. The final absolute parameters of V486 Car as determined by our analysis, together with their errors, are listed in Table \ref{table:abs_par}. These photometric LC modellings indicate V486 Car to be an shallow-contact system with poor thermal contact. Accordingly, the components have relatively large radii, and hence luminosities, according to their masses.

\begin{table}
\centering
\caption{Absolute parameters of V486 Car.}
\label{table:abs_par}
\begin{tabular}{lrr}
\hline
Parameter		                    & Value      & Uncertainty\\
\hline
$a$~(R$_{\odot}$)	                & 5.91        & 0.01      \\		
$M_1$~(M$_{\odot}$) 	            & 2.06        & 0.11	  \\		
$M_2$~(M$_{\odot}$) 	            & 0.35        & 0.04      \\		
$R_1$~(R$_{\odot}$) 	            & 3.20        & 0.02	  \\		
$R_2$~(R$_{\odot}$)	                & 1.48        & 0.01      \\		
$\log g_1$	                        & 3.74        & 0.03	  \\		
$\log g_2$ 	                        & 3.14        & 0.02	  \\		
$T_1$ (K) 	                        & 10000       & 500	      \\		
$T_2$ (K) 	                        & 6226        & 150	      \\		
$L_1$~(L$_{\odot}$)                 & 92.05       & 19.6	  \\		
$L_2$~(L$_{\odot}$)            	    & 2.97        & 0.33	  \\
$M_{\rm bol,~1}$ (mag) 	            & $-0.172$    & 0.231	  \\		
$M_{\rm bol,~2}$ (mag) 	            & 3.560       & 0.119	  \\		
$M_{\rm bol,~system}$ (mag) 	    & $-0.206$    & 0.260     \\	
$BC_{\rm TESS,~1}$ (mag) 	        & $-0.33$     & 0.01	  \\		
$BC_{\rm TESS,~2}$ (mag) 	        & 0.44        & 0.01	  \\
$TESS$ (mag) 	                    & 6.23        & 0.01      \\
$A_{\rm TESS}$ (mag)                & 0.10        & 0.01      \\
$M_{\rm TESS,~system}$ (mag) 	    & 0.089       & 0.126     \\	
$d$ (pc)                    	    & 162         & 12        \\	
\hline
\end{tabular}
\end{table}

A check on the accuracy of the absolute parameters, given in Table \ref{table:abs_par}, is to compare the photometric parallax with the trigonometric parallax of Gaia DR3. The photometric parallax can be computed from the following equations given by \cite{Budding_Demircan_2007}:
\begin{equation}
\log \pi  = 7.454 - \log R - 0.2 \text{V} - 2 F'_\text{V} \\
\label{eq:photo_paralax}
\end{equation}
{where $F'_\text{V}$ (flux scale) is equal to 0.25 $\times$ the logarithm of the surface flux in the V band, and is specified by}
\begin{equation}
F'_\text{V} = \log T_{e} + 0.1 BC
\label{eq:flux_scale}
\end{equation}

Applying Eq.~\ref{eq:photo_paralax} and Eq.~\ref{eq:flux_scale} directly to the components of V486 Car, with the $V_0$ magnitudes from Table~\ref{tab:WD_colors}, the $R$ values from Table \ref{table:abs_par} and $T_{e}$ values from V-band LC solution from Table~\ref{Tab:WD+MC BV&Hp Model}, and the BC values from \cite{Eker_etal_2018}, we obtain $\log \pi_1 = -2.22$, and $\log \pi_1 = -2.23$.  These parallaxes are in close agreement with the value cited above from the Gaia-DR3 ($\log \pi = 2.21$).

The consistency of the physical parameters derived for the components of the V486~Car system, as well as the system distance, can be examined through its spectral energy distribution (SED) constructed over a wide wavelength range. To this end, the SED of the system was modelled using the method originally introduced by \citet{2022AcA....72..195B} and subsequently refined by \citet{2023MNRAS.523.2440E}.

In this modeling procedure, the fluxes of the individual components were reconstructed using  LTE grid models with the new ODF files of \citet{2003IAUS..210P.A20C}. These take into account the effective temperatures and  surface gravities of the components. The synthetic spectra were generated with the \textsc{ATLAS9} and \textsc{SYNTHE} codes, while interstellar extinction was treated as the only free parameter in the SED fitting.

\begin{figure}
    \centering
    \includegraphics[width=\columnwidth]{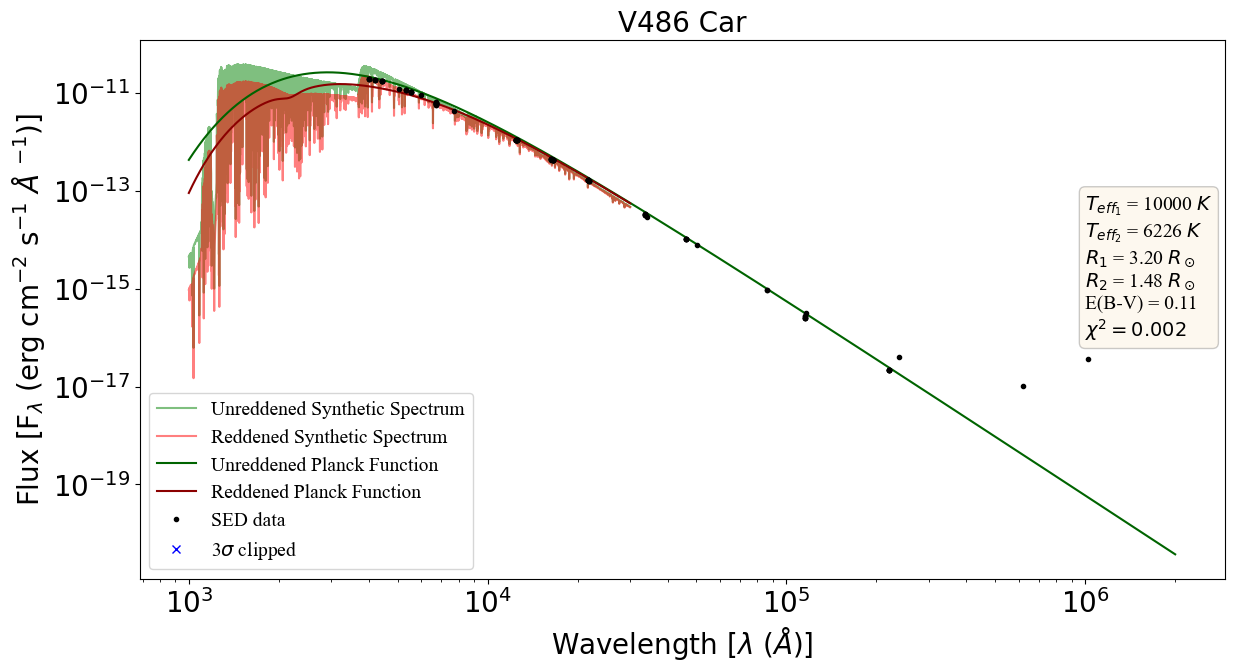}
    \caption{Spectral energy distribution of the V486~Car system.
    Filled circles represent the observed broadband fluxes compiled from the literature. The solid red curves denote the best-fitting reddened composite SED models based on the Planck function and synthetic spectra of the primary and secondary components, constructed using the derived effective temperatures and surface gravities. The solid green curves correspond to the same models without interstellar extinction. The reddened models include interstellar extinction with a best-fitting value of $E(B-V)=0.11\pm0.01~\mathrm{mag}$. }
    \label{fig:sed}
\end{figure}

The best-fitting model to reproduce the observed SED of the system yields an interstellar reddening of $E(B-V)=0.11\pm0.01~\mathrm{mag}$. Fig.~\ref{fig:sed} demonstrates the agreement between the observed SED and the synthetic models constructed using the derived physical properties of the system components.

\section{Discussion  and conclusions}
\label{discussion}

Our adopted  absolute parameters for the close binary system V486 Car, as given in Table~\ref{table:abs_par}, and illustrated, in part, in Fig~\ref{fig:wd_hotspot_model}. The parameter values bring up the issue, evident from compilations of data like those of \cite{Kirk_2016},  of non-conformity with standard single dwarf star evolution models \citep{Hilditch_2001}.   This particularly concerns stars being over-luminous for their mass. 

To be sure, over-luminous secondary stars are a regular arrangement in classical Algols, testifying to the well-known model of interactive binary evolution with mass transfer by the process of Roche Lobe Overflow (RLOF).   The root cause of this over-luminosity appears to be that the mass-losing star has moved to a more photo-emissive phase in its evolution. For classical Algols the `subgiant' secondary is frequently powered by  a helium-burning shell in its core region  (`Case B'), while the primary's luminosity may be supplemented by that of in-falling  transferred matter.  

Using Table~9.2 in \cite{Budding_2022},  and the parameters given in Table~\ref{table:abs_par}, we deduce that the present primary star of V486 Car is $\sim$92 solar luminosities greater than the Sun, or about 0.25 magnitudes brighter than a single Main Sequence star of A0V type. The secondary is 2.86 times more luminous than the Sun, so its low mass $\sim0.4$ M$_{\odot}$ indicates that this is an evolved object, or perhaps the recipient of extensive energy transfer. Interactive evolution points to the original mass having been greater than  $\sim$2 M$_{\odot}$, most of which would has been transferred to the present primary.  

This present primary continues in its over-energized state: at some point it may return matter back to the present secondary if not already doing that. Stars in interactive binary systems  thus show strikingly different patterns of behaviour than single stars: the Vogt-Russell theorem being, in general, incompatible with the Roche Lobe filling constraint \citep{Kuiper_1941}. We could then expect the system to be unstable on the Kelvin timescale, in effect conforming to the back-and-forth condition of thermal relaxation oscillations (TRO, \citeauthor{Lucy_1968a}, \citeyear{Lucy_1968a, Lucy_1968b},  \citeauthor{Hazlehurst_1973},
\citeyear{Hazlehurst_1973}; \citeauthor{Flannery_1976}, \citeyear{Flannery_1976}).

Evidence about the low mass secondary star comes from the LC analysis -- even the initial LC that shows a difference in depths of the two minima, although the amplitude is small.  More detailed photometric study,  illustrated in Figs.~\ref{fig:wd_hotspot_model} \& \ref{fig:wf_model},
confirm this . Further general support comes from the  $K_1$, $K_2$ normalized RMS map obtained  from the  {\sc korel} program and shown in Fig~\ref{fig:chi_map}.  The LC and RV fittings agree  on the near-contact status of the binary, although the adopted value for the mass ratio results in both stars having fractional radii that are too large for the contact lobes (see Tables~4.2 and 4.3 in \citeauthor{Budding_2022}, \citeyear{Budding_2022}) (Tables 4 \&  5 ibid.).

This over-contact configuration of V486 Car is not typical for well-known examples, such as  W UMa or $\epsilon$ CrA. where orbital periods are generally quite shorter and surface temperatures cooler. V486 Car would be positioned at the extremity  of the distribution shown as Fig. 6.7 by 
\cite{Hilditch_2001}. More typical contact systems are thought to progress to an eventual merger of the binary components \citep{Webbink_1976, Webbink_2003, Webbink_2006}.
However, the one established merger event,  V1309 Sco, with its relatively long orbital period ($\sim$1.44\,d) and low mass ratio, gives reason to be cautious regarding expectations of the  evolution of near-contact  binary stars \citep{Tylenda_2011}.

While an over-contact condition might be formally  calculable, the consequences of barotropically maintaining two very different effective temperatures on the same common photosphere appear non-physical.  This over-contact model becomes more plausible if there is a slight error  in the small numerator or denominator on the right of Eqn~\ref{eq:R_J}, or else in the spectroscopic mass ratio $q$ in Table~\ref{tab:orbit_pars}.  This option is not at odds  with the probability contours of Fig.~\ref{fig:chi_map}. Thus, if we set $q = 0.15$,  we obtain the reasonable arrangement of the secondary filling its Roche Lobe, with a slightly detached primary (Table~\ref{tab:contactqr}), i.e.\  the {\em semi-detached }  configuration that forms a naturally required stage in the TRO theory. A lower value of $q$, implying lower $k^2$,
also reduces the temperature disparity following from Eqn~\ref{eq:R_J}.

Such an arrangement fits easily into the very close stage of classical Algol evolution, and has particular relevance to the origin of R CMa-type binaries \citep{Bakis_2024}.  However.
it does not, {\it prima facie},  align with the sense of the O -- C variation shown in the middle panel of Fig.~\ref{fig:o-c_all}, that suggests matter transferred from the present primary to the secondary (period shortening).  Of course, there may be other factors at play in the data-sparse Fig.~\ref{fig:o-c_all}. In Section \ref{O_Connell} it was argued
that the variable O'Connell effect would interfere with interpretations of times of minima, as, to some extent, would the low-amplitude, $\sim$10 d period `jitter' discovered in this study. The recent 
work  of \cite{Fabry_2025} supports  the 
general explanation of the O'Connell asymmetry in
in terms of mass transfer effects
\citep{Bakis_2024}.

So, the foregoing interpretations, while feasible, appears ambivalent and leave our understanding of V486 Car unresolved.  The situation calls for further, more comprehensive  and precise  observations and modelling in order to gain a fuller and firmer appreciation of the structure and evolution of close binary stars.

\begin{table}
\begin{center}
\caption{Mass-ratio $q$ and contact relative radii $r_{1,2}$.
\label{tab:contactqr}}
\footnotesize
\begin{tabular}{llll}
  \hline 
\multicolumn{1}{l}{Parameter}  & \multicolumn{3}{c}{Value}  \\
\hline
$q$    &    0.2   & 0.175    &  0.15  \\  
$r_1$  &   0.5297 &  0.5420  & 0.5573  \\
$r_2$  &   0.2422 &  0.2328  & 0.2223   \\
\hline 
\end{tabular}
\end{center}
\end{table}

\section{Acknowledgments}
Generous allocations of time on the 1m McLennan Telescope and {\sc hercules} spectrograph at the Mt John University Observatory in support of the Southern Binaries Programme have been made available through its TAC and supported by its  Director, Dr.\ K.\ Pollard and previous Director, Prof.\ J.\ B.\ Hearnshaw. Useful help at the telescope was provided by the MJUO management (N.\ Frost and previously A.\ Gilmore \& P.\ Kilmartin). Considerable assistance with the use and development of the {\sc hrsp} software was given by its author Dr.\ J.\ Skuljan, and very helpful work with initial data reduction was carried out by R.\ J.\ Butland.   We thank the University of Queensland for the collaboration software. This paper includes data collected by the TESS mission and obtained from the MAST data archive at the Space Telescope Science Institute (STScI). STScI is operated by the Association of Universities for Research in Astronomy, Inc., under NASA contract NAS 5–26555.   Funding for the TESS mission is provided by the NASA's Science Mission Directorate. This research has made use of the SIMBAD database, operated at CDS, Strasbourg, France, and of NASA's Astrophysics Data System Bibliographic Services. We also made use of data from the European Space Agency (ESA) mission {\it Gaia} (\url{https://www.cosmos.esa.int/gaia}), processed by the {\it Gaia} Data Processing and Analysis Consortium (DPAC, \url{https://www.cosmos.esa.int/web/gaia/dpac/consortium}). Funding for the DPAC has been provided by national institutions, in particular the institutions participating in the {\it Gaia} Multilateral Agreement.

\section*{Data Availability}

The majority of data included in this article are available as listed in the paper. The TESS data are available online from the MAST repository (\url{https://mast.stsci.edu/portal/Mashup/Clients/Mast/Portal.html}.


\bibliographystyle{mnras}
\bibliography{references} 



\clearpage
\appendix
\section{Table of Times of Eclipses}
\label{secA}

\begin{center}
\captionof{table}{Times of eclipses for V486 Car. `Pri' indicates a primary eclipse, and `sec' a secondary one. BVI refer to the selected Johnson filters in the Congarinni Observatory data.}
\label{table:ToMs}

\begin{tabular}{llcc}
\hline
  Time         & Error & Type & Source  \\[1pt] 
 [BJD-2400000] & [d]   & pri$/$sec  &  \\
\hline
48230.7024 & 0.0013 & pri & Hipparcos \\
48778.7395 & 0.0010 & pri & Hipparcos \\
56063.9958 & 0.0050 & pri & I \\
60703.1684 & 0.0150 & pri & V \\
60770.9789 & 0.0080 & pri & V \\
60770.9892 & 0.0100 & pri & B \\
48231.2739 & 0.0016 & sec & Hipparcos \\
48779.3082 & 0.0015 & sec & Hipparcos \\
56000.0308 & 0.0050 & sec & I \\
60729.9681 & 0.0170 & sec & B,V \\
\hline
\end{tabular}
\end{center}

\section{Results of the WD+MC fitting to the TESS light curves}

\begin{center}
\captionof{table}{Results of the WD+MC hot spot model fitting to the TESS light curves of V486 Car.}
\label{Tab:WD+MC_TESS_hotspot_Model}
\begin{tabular}{lccccc}
\hline
Parameter   & Sector 09 & Sector 10 & Sector 63 & Sector 64 & Sector 65 \\
\hline
$\Delta \phi$               & $0.0087$ $\pm0.0003$     & $0.0085$ $\pm0.0004$     & $0.0080$ $\pm0.0004$     & $0.0078$ $\pm0.0005$     & $0.0078$ $\pm0.0004$	\\
$i$ ($\degr$)               & 51.119 $\pm0.346$      & 51.431 $\pm0.205$      & 51.365 $\pm0.197$      & 51.543 $\pm0.216$      & 51.374 $\pm0.206$	\\
$T_1$ (K)                   & 10000 (fixed)         & 10000 (fixed)         & 10000 (fixed)         & 10000 (fixed)         & 10000 (fixed) \\
$T_2$ (K)                   & 6247 $\pm91$ 	        & 6184 $\pm88$           & 6247 $\pm86$           & 6157 $\pm99$           & 6280 $\pm91$	\\
$\Omega_1 = \Omega_2$       & 2.1323 $\pm0.0042$ 	& 2.1374 $\pm0.0019$     & 2.1383 $\pm0.0024$     & 2.1373 $\pm0.0023$     & 2.1383 $\pm0.0023$	\\
$q=M_2/M_1$ 	            & 0.174 (fixed)         & 0.174 (fixed)         & 0.174 (fixed)         & 0.174 (fixed)         & 0.174 (fixed) \\
Fill-out factor             & 0.31                  & 0.26                  & 0.25                  & 0.26                  & 0.25 \\
$r_1$ (volume)	            & 0.544 $\pm0.002$      & 0.542 $\pm0.001$      & 0.542 $\pm0.001$      &  0.542 $\pm0.001$      &  0.542 $\pm0.001$	\\
$r_2$ (volume)	            & 0.251 $\pm0.008$       & 0.250 $\pm0.004$ 	    & 0.250 $\pm0.004$	    &	 0.250 $\pm0.004$    & 0.250 $\pm0.004$	\\
$L_1/(L_1+L_2)$             & 0.93 $\pm0.03$ 	    & 0.94 $\pm0.03$         & 0.93 $\pm0.03$         & 0.94 $\pm0.03$         & 0.93 $\pm0.03$	\\
$L_2/(L_1+L_2)$             & 0.07 $\pm0.01$     & 0.06 $\pm0.01$         & 0.07 $\pm0.01$         & 0.06 $\pm0.01$    & 0.07$\pm0.01$	\\
\hline
Spot parameters \\
$\beta$ ($\degr$)           & 90 (fixed)            & 90 (fixed)            & 90 (fixed)            & 90 (fixed)            & 90 (fixed) \\
$\lambda$ ($\degr$)         & 142 $\pm2$            & 143 $\pm2$            & 140 $\pm2$            & 143 $\pm3$            & 141 $\pm2$   \\
$\gamma$ ($\degr$)          & 52 $\pm2$             & 42 $\pm3$             & 39 $\pm2$             & 41 $\pm3$             & 38 $\pm3$ \\
$\kappa$                    & 1.243 $\pm0.026$      & 1.342 $\pm0.059$      & 1.353 $\pm0.063$      & 1.338 $\pm0.067$      & 1.356 $\pm0.075$  \\
\hline
$\chi^{2}_{\rm red}$        &	1.23	            &	1.69	            &	1.14	            &	0.96	            &	1.08 \\
$\nu$      	                &	16150	            &	15522	            &	18337	            &	18833	            &	18350	\\
$\Delta l$ 	                &	0.0015	            &	0.0015	            &	0.0015	            &	0.0015	            &	0.0015 \\
\hline												
\end{tabular}
\end{center}

\clearpage
\begin{center}
\captionof{table}{Results of the WD+MC cool spot model fitting to the TESS light curves of V486 Car.}
\label{Tab:WD+MC_TESS_coolspot_Model}
\begin{tabular}{lccccc}
\hline
Parameter   & Sector 09 & Sector 10 & Sector 63 & Sector 64 & Sector 65 \\
\hline
$\Delta \phi$   & $0.0014\pm0.0005$ & $-0.0003$ $\pm0.0004$ & $-0.0008$ $\pm0.0005$ & $-0.0008$ $\pm0.0004$ & $-0.0008$ $\pm0.0005$	\\
$i$ ($\degr$)   & 52.380 $\pm0.374$ & 52.174 $\pm0.358$ & 52.496 $\pm0.328$ & 52.480 $\pm0.341$ & 52.553 $\pm0.352$	\\
$T_1$ (K)   & 10000 (fixed) & 10000 (fixed) & 10000 (fixed) & 10000 (fixed) & 10000 (fixed) \\
$T_2$ (K)   & 7510 $\pm122$ & 7464 $\pm118$ & 7334 $\pm105$ & 7372 $\pm108$ & 7402 $\pm105$	\\
$\Omega_1 = \Omega_2$ & 2.1191 $\pm0.0036$ & 2.1217 $\pm0.0032$ & 2.1193 $\pm0.0035$ & 2.1190 $\pm0.0035$ & 2.1196 $\pm0.0034$	\\
$q=M_2/M_1$ & 0.174 (fixed) & 0.174 (fixed) & 0.174 (fixed) & 0.174 (fixed) & 0.174 (fixed) \\
Fill-out factor & 0.43 & 0.40 & 0.43 & 0.43 & 0.42 \\
$r_1$ (volume) & 0.549 $\pm0.002$ & 0.548 $\pm0.002$ & 0.549 $\pm0.002$ & 0.549 $\pm0.002$ &  0.549 $\pm0.002$	\\
$r_2$ (volume)	& 0.256 $\pm0.001$ & 0.255 $\pm0.001$ & 0.256 $\pm0.001$ & 0.256 $\pm0.001$    & 0.256 $\pm0.001$	\\
$L_1/(L_1+L_2)$ & 0.89 $\pm0.03$ & 0.89 $\pm0.03$ & 0.90 $\pm0.03$ & 0.90 $\pm0.03$ & 0.90 $\pm0.03$	\\
$L_2/(L_1+L_2)$ & 0.11 $\pm0.01$ & 0.11 $\pm0.01$ & 0.10 $\pm0.01$ & 0.10 $\pm0.01$ & 0.10 $\pm0.01$	\\
$l_3$   & 0.09 $\pm0.02$ & 0.08 $\pm0.02$ & 0.10 $\pm0.02$ & 0.10 $\pm0.02$ & 0.10 $\pm0.02$ \\ 
\hline
Spot parameters \\
$\beta$ ($\degr$) & 100 $\pm14$ & 102 $\pm11$ & 97 $\pm14$ & 99 $\pm14$ & 100 $\pm14$ \\
$\lambda$ ($\degr$) & 296 $\pm9$ & 294 $\pm7$ & 296 $\pm8$ & 296 $\pm8$ & 295 $\pm8$   \\
$\gamma$ ($\degr$) & 33 $\pm10$ & 33 $\pm7$ & 33 $\pm10$ & 33 $\pm10$ & 36 $\pm10$ \\
$\kappa$ & 0.584 $\pm0.050$ & 0.550 $\pm0.069$ & 0.619 $\pm0.051$ & 0.610 $\pm0.053$ & 0.666 $\pm0.046$  \\
\hline
$\chi^{2}_{\rm red}$ & 0.87 &	1.33 &	0.94   &	0.67 & 0.84 \\
$\nu$ &	16150	&	15522 &	18337 &	18833 &	18350	\\
$\Delta l$ & 0.0015 & 0.0015 & 0.0015 &	0.0015 & 0.0015 \\
\hline												
\end{tabular}
\end{center}

\clearpage
\section{Spectroscopy}

\begin{center}
\captionof{table}{Log of spectroscopic observations for V486 Car.}
\label{tab:observing_log}
\scalebox{0.91}{
\begin{tabular}{lccccc}
  \hline 
Image	&	Date	&	BJD	&	Phase	&	Exp. Time  &	S/N	\\
        & mm/dd/yr & --2450000     &           & (s)       & ratio \\
\hline
w5542024	&	12/11/2010	&	5542.0568	&	0.8454	&	834	&	65	\\
w5542026	&	12/11/2010	&	5542.0697	&	0.8573	&	1500	&	75	\\
w5545007	&	12/14/2010	&	5544.9033	&	0.4476	&	1407	&	64	\\
w5545009	&	12/14/2010	&	5544.9257	&	0.4682	&	1425	&	71	\\
w5545013	&	12/14/2010	&	5544.9756	&	0.5138	&	1000	&	63	\\
w5545019	&	12/14/2010	&	5545.0381	&	0.5709	&	2000	&	71	\\
w5796008	&	8/22/2011	&	5795.8370	&	0.8446	&	849	&	65	\\
w5796015	&	8/22/2011	&	5795.8804	&	0.8843	&	1000	&	75	\\
w5796019	&	8/22/2011	&	5795.9094	&	0.9108	&	1000	&	68	\\
w5796023	&	8/22/2011	&	5795.9426	&	0.9411	&	1235	&	68	\\
w5796030	&	8/22/2011	&	5795.9890	&	0.9835	&	1380	&	69	\\
w5796052	&	8/22/2011	&	5796.1918	&	0.1689	&	775	&	60	\\
w5796056	&	8/22/2011	&	5796.2278	&	0.2019	&	655	&	65	\\
w5798032	&	8/24/2011	&	5798.0249	&	0.8447	&	1418	&	70	\\
w5798042	&	8/24/2011	&	5798.0843	&	0.8991	&	1118	&	65	\\
w5798050	&	8/24/2011	&	5798.1580	&	0.9665	&	972	&	70	\\
w5798052	&	8/24/2011	&	5798.1865	&	0.9925	&	762	&	71	\\
w5876022	&	11/10/2011	&	5875.9920	&	0.1202	&	903	&	60	\\
w5876024	&	11/10/2011	&	5876.0042	&	0.1314	&	900	&	67	\\
w5876034	&	11/10/2011	&	5876.0929	&	0.2124	&	1289	&	62	\\
w5876036	&	11/10/2011	&	5876.1128	&	0.2307	&	1000	&	67	\\
w5882009	&	11/16/2011	&	5882.1001	&	0.7041	&	1000	&	68	\\
w5882015	&	11/16/2011	&	5882.1595	&	0.7584	&	1200	&	66	\\
w5882017	&	11/16/2011	&	5882.1791	&	0.7763	&	1370	&	68	\\
w5883017	&	11/17/2011	&	5882.9192	&	0.4528	&	1000	&	60	\\
w5883022	&	11/17/2011	&	5882.9794	&	0.5079	&	1500	&	68	\\
w6255009	&	11/23/2012	&	6255.0606	&	0.6547	&	999	&	62	\\
w6258008	&	11/26/2012	&	6257.9163	&	0.2653	&	1664	&	65	\\
w6258017	&	11/26/2012	&	6257.9884	&	0.3312	&	1382	&	67	\\
w6668036	&	1/10/2014	&	6668.0672	&	0.2145	&	577	&	52	\\
w6671014	&	1/13/2014	&	6670.9319	&	0.8333	&	900	&	60	\\
w6672017	&	1/14/2014	&	6671.9238	&	0.7400	&	819	&	56	\\
w6674041	&	1/16/2014	&	6674.0701	&	0.7022	&	1266	&	78	\\
w6993035	&	12/1/2014	&	6993.1420	&	0.3893	&	1057	&	75	\\
w6993037	&	12/1/2014	&	6993.1550	&	0.4012	&	872	&	74	\\
\hline 
\end{tabular}}
\end{center}

\begin{table}
\begin{center}
\caption{Identified spectral lines for V486 Car based on comparison with the ILLSS Catalogue \citep{Coluzzi_1993, Coluzzi_1999}. The lines are considerably broadened; to the scale of the orbital motion $\sim$300 km s$^{-1}$ and must therefore include blends of both components.}
\label{tab:spectral_features}
\scalebox{0.78}{
\begin{tabular}{lclll}
  \hline 
\multicolumn{1}{l}{Species}  & \multicolumn{1}{c}{Order no.} &
\multicolumn{1}{l}{Adopted $\lambda$} & \multicolumn{1}{l}{Comment}  \\
\hline
H$_{\alpha}$        & 87        &  6562.82      &   strong, blended \\
Fe II          & 88   & 6456.38 & weak \& blended  \\
Fe II               &  89       &  6375.87 &   visible, telluric intrusions    \\
Si II               & 89        & 6371.36             &   visible \\
Si II               & 90        & 6347.10             &  rel. strong   \\
Fe I  &91         & 6267.85     &             \\
Fe II & 91   & 6247.56          & weak   \\
Fe II               & 91        & 6238.38          &  weak       \\
Fe I  & 92            & 6157.41, 6157.73        & broad \& blended \\
Si II  & 95   & 5978.97    &  telluric intrusions \\
Na I  & 96    &  5889.95, 5895.92   &  strong and narrow \\
He I   & 97     &  5875.85   &  He I triplet (av.\ $\lambda$) Aa \\
Fe I   & 102           & 5658.83    & \\ 
Mg I triplet lines  & 103 & 5528 & weak \& blended \\
Fe II  & 103   & 5527.35    & weak \& blended \\
Fe II   & 103   & 5534.86   & weak \& blended  \\
Fe II   & 104   & 5465.93  & weak \& blended  \\
Fe I   & 104  & 5446.51, 5455.61  & weak \& blended  \\
Fe I  & 107   & 5339.92, 5341.02  & visible \& blended \\
Fe II   & 107   & 5315.08, 5316.21, 5318.05 & strong \& blended \\
Fe II, Fe I      & 107   & 5325.54, 5328.93      & visible \& blended \\
Ca I, Fe II   & 108         &  5264               & weak \& blended     \\
Fe II   & 108   & 5275.99, 5284.09     & visible \& blended \\
Ti II, Fe I   & 109   & 5226.54, 5227.19     & strong \& blended \\
Fe I      & 109   & 5234.62   & strong \\
Mg I      & 110   & 5183.6042    & strong \\
Mg I, Fe I      & 110   & 5172.68, 5173.12     & strong \& blended \\
Mg I, Fe II     & 110   & 5167.32, 5169.80     & strong \& blended \\
Si II      & 112, 113      & 5056.02   &    weak     \\
Fe II    & 113           &  5018.434     &  strong  \\
Fe I      & 115   & 4957    & weak \\
Mg I      & 115   & 4923.2     & very weak \& blended \\
He I         & 115      &  4921.92  & detectable                   \\
Fe I, Fe II  & 115 & 4920.509, 4923.921 & strong \& blended \\
Ti I  & 115    & 4919.867    &  Ti I and Fe I blend   \\  
Fe             & 116      & 4891.496     &      weak          \\
H$_{\beta}$   & 117           &  4861.3  &  strong, blended \\
Mn I          & 118           &  4823.516       & weak     \\
N II      & 121           &  4704.33                 & weak      \\
Mn I      & 121           &  4701.16                 &  blend        \\
Fe II, Cr II & 123  & 4629, 4634 & weak \& blended \\
Cr II     & 123   & 4618.81     & weak \& blended \\
Cr II     & 124   & 4589.217    &   weak \\
Cr II     & 124   & 4588.217     & weak   \\
Fe II     & 124   & 4583.829     & strong \\
Fe II     & 124           &  4582.835    &   \\
Mg I     & 124           &  4571.096     &   blend with Ti II \\
Ti II     & 124   & 4571.97     & blend with Mg I \\
Ti II     & 124   & 4563, 4571  & strong \& blended \\
Ti II     & 125   & 4571.97     & blend with Mg I \\
Ti II     & 125   & 4563.761    & weak \\
Fe II, Cr II     & 125   & 4555.89, 4558.64     & weak \& blended \\
Fe II, Ti I    & 125        & 4549.467, 4549.622               & strong \& blended         \\
Ti II     & 125   & 4533.966     & weak \\
\hline
\end{tabular}}
\end{center}
\end{table}

\bsp
\label{lastpage}
\end{document}